\documentstyle[aps,preprint,epsf]{revtex}

\begin{document}
\draft

\title{Phase Diagram of Amorphous Solid Water: Low-Density,
High-Density, and Very-High-Density Amorphous Ices}

\author{Nicolas Giovambattista$^1$, H. Eugene Stanley$^2$ and
 Francesco Sciortino$^3$}

\address{
$^1$ Department of Chemical Engineering,\\
Princeton University, Princeton, NJ 08544-5263 USA\\
$^2$Center for Polymer Studies and Department of Physics\\
Boston University, Boston, MA 02215 USA \\
$^3$ Dipartimento di Fisica and INFM Udr and 
SOFT: Complex Dynamics in Structured Systems\\
Universita' di Roma ``La Sapienza'' -- 
Piazzale Aldo Moro 2, I-00185, Roma, Italy
}

\date{21 February 2005}

\maketitle

\begin{abstract}

We describe the phase diagram of amorphous solid water by performing
 molecular dynamics simulations using the simple point charge (SPC/E)
 model. Our simulations follow
 different paths in the phase diagram: isothermal compression/decompression,
 isochoric cooling/heating and isobaric cooling/heating.
 We are able to identify low-density
 amorphous (LDA), high-density amorphous (HDA), and very-high density
 amorphous (VHDA) ices. The density $\rho$ of these glasses at different
 pressure $P$ and temperature $T$ agree well
 with experimental values. We also study the radial
 distribution functions of glassy water.  
In agreement with experiments, we find
that LDA, HDA and VHDA are characterized by a tetrahedral hydrogen-bonded network and that, as compared to LDA, HDA has an extra
interstitial molecule between the first and second shell. VHDA appears to have  two such extra molecules.  We obtain VHDA by isobaric  heating of HDA, as in experiment. We also find that ``other forms'' of glassy water can be obtained upon isobaric heating of LDA, as well as amorphous ices formed 
during the transformation of LDA to HDA.
 We argue that these other forms of amorphous ices, as well as VHDA,  are not
altogether new glasses but rather are the result of aging  induced by heating.  Samples of HDA and VHDA with different densities are recovered at normal $P$, showing that there is a continuum of glasses. Furthermore, the two ranges of densities of recovered HDA and recovered VHDA overlap at
ambient $P$.  Our simulations are consistent with the possibility  of HDA$\rightarrow$LDA and
 VHDA$\rightarrow$LDA transformations, reproducing the experimental
 findings. We do not observe a VHDA$\rightarrow$HDA transformation, and our final
 phase diagram of glassy water together with equilibrium liquid data suggests
 that for the SPC/E model the VHDA$\rightarrow$HDA 
transformation cannot be observed with the present heating rates accessible in simulations. 
Finally, we discuss the consequences of our findings for the understanding of the transformation between the different amorphous ices and the two
hypothesized phases of liquid water.

\end{abstract}



\section{Introduction}

Liquids can transform into glass if they are cooled fast enough to avoid
 crystallization. Water is not an exception, although vitrifying liquid water
 requires very high quenching rates ($\approx 10^6$~K/s)\cite{HGW}
--- the resulting glass is called hyperquenched glassy water (HGW).
Actually, water can exist in more than one
amorphous form, a property called ``polyamorphism'' 
 \cite{PabloReview,AngellReview,geneNature}.
Uniaxial
compression of ice-$I_h$\cite{mishima84} or
ice-$I_c$\cite{floriano,joharyIc} at temperature $T=77$~K to pressures 
$P \gtrsim 1$~GPa produces a disordered high density material named
 high-density amorphous ice (HDA).  If HDA is recovered at $T=77$~K and
 $P=1$~bar and then heated isobarically, it transforms irreversibly at $T
 \approx 125$~K to a disordered low-density material named 
 low-density amorphous ice (LDA) \cite{mishima84,mishimaNature85}.  
Both HGW and LDA transform to HDA when compressed at $T=77$~K at
$P=0.6$~GPa\cite{mishimaNature85,mishimaVisual}.  X-ray and
neutron diffraction measurements suggest that LDA is structurally identical
 to HGW \cite{bellisent} and, although small differences have been
 found\cite{waterAB,klugPRL,TseKlug}, the common view is that HGW
 and LDA, are the same material
 \cite{PabloReview,geneNature,genePhysToday}. 

The LDA-HDA transformation is reversible if it is performed above $T\approx 130$~K\cite{mishimaJCP}. Furthermore, because the LDA-HDA transformation is very sharp and shows hysteresis, it has been suggested that it is a first-order transition \cite{mishimaJCP,mishimaLL}.
It has also been proposed\cite{poolepre} to interpret the LDA-HDA transition
 as the glass counterpart of the first order transition between two liquids
 of different structure, the liquid-liquid phase transition scenario
\cite{poole}.
 In this view, LDA is the glass associated with a low-density
 liquid (LDL) while HDA is the glass associated with
 a high-density liquid (HDL)
\cite{mishimaLL,poolepre,mishimaJCP2001,PELWater,reichert}.  

Computer simulations have been able to reproduce many of the experimental
results.  More than 15 years ago, simulations using the TIP4P model found a
transformation of ice-$I_h$ into HDA when 
compressed at $T=80$~K and a transformation of HDA to LDA upon isobaric heating at zero $P$ \cite{tse}. The LDA-HDA reversible
transformation is also found in simulations using the ST2\cite{poole} and
TIP4P\cite{poolepre} models. 
Computer simulations in the liquid phase also show a LDL-HDL first-order
transition\cite{poole,harrington,masako,meyer}. 

Very recently, a glass denser than HDA was identified in experiments, and
named very-high density amorphous (VHDA) ice\cite{loerting}.
 VHDA is obtained by isobaric heating at high $P$ of HDA from $T=77$~K
 to $T\approx 165$~K.  Two samples of HDA, one heated isobarically 
at $P=1.1$~GPa and the other at $P=1.9$~GPa, both relax at
$T=77$~K and $P=1$~bar to the same structure with a density of
$\rho = 1.25 \pm 0.01$~g/cm$^3$, i.e., $\approx 9\%$ denser than HDA 
($\rho= 1.15$~g/cm$^3$) and $\approx 40\%$ denser than LDA
($\rho=0.94$~g/cm$^3$). It
appears that the properties of HDL are closer to VHDA
than to HDA \cite{finneyHDAVHDA,french}.  
In recent reports \cite{parrinello,ourVHDA}, it has been shown that
VHDA can be obtained in computer simulations  
by following the same procedure as in experiments. Further, we argued
\cite{ourVHDA} that
 VHDA may not be a new glass different than HDA, but rather VHDA (and not HDA) 
results from partial annealing of the ``more metastable'' 
HDA structure. Further, our simulations also showed that VHDA
is the glass obtained by isobaric quenching of the equilibrium liquid at high
$P$. 

It is an open question whether all forms of glassy water can be classified
 into two amorphous forms, which we call LDA and VHDA.  While this might be
 the case for glasses obtained by cooling liquid water or compressing 
 ice, there are many different experimental techniques
to obtain glassy water. For instance, amorphous ice can be also obtained by
exposing crystalline ice to radiations such as  electrons\cite{GG8}, ultraviolet photons\cite{GG10}, and ion bombardment\cite{GG11}. 
 The situation is more complicated when considering the effect of aging in the glassy state. Recent works\cite{science,urquidi2,mishimaSuzuki}
suggest that aging takes place
 on the experimental time scale when HDA is kept at ambient $P$ and
$T$ is below the HDA-LDA transformation $T$. The $\rho$ and structure of the sample changes with the 
 annealing $T$ and annealing time. 
 

In this work we rebuild the phase diagram of amorphous water on the basis of
 simulations using the SPC/E model.  Our simulations consist mainly 
of isobaric cooling/heating and isothermal compression/decompression  cycles. In section~\ref{simulations}, we describe briefly the simulation details. In section~\ref{LDAtoHDA}, we show that compression of LDA leads to HDA. Isobaric heating at different $P$s is reported in section~\ref{HDAtoVHDA}, where the HDA$\rightarrow$VHDA transformation
 is also described. In section~\ref{recoveredGLASSES}, we describe the results of recovering HDA and VHDA at normal $P$. The structures of all glasses are
studied in section~\ref{rdfs}.  The HDA$\rightarrow$LDA, VHDA$\rightarrow$LDA and VHDA$\rightarrow$HDA transitions are
investigated in section~\ref{x-x}.   Finally,
 in section~\ref{summary} we discuss our results and recent experiments in the context of a possible
 first order transition line  separating two amorphous ices at low $T$, and two liquid phases at high  $T$.

\section{Simulations}
\label{simulations}

We perform molecular dynamics (MD) simulations using the extended simple 
point charge (SPC/E) model of water\cite{berensen}.  This model has been
extensively used to study the thermodynamics
\cite{francislong,pooleSPC/E} and dynamics\cite{francescoSPCE2,ourSHD}
of liquid water. Thermodynamic and dynamical properties of SPC/E water  are
well known, and are consistent with experimental facts. The SPC/E model
reproduces the thermodynamic anomalies characterizing water (e.g. it shows a
maximum in $\rho$ \cite{francescoSPCE1}).  This model has also been used to
study glassy water, and is able to produce glassy states corresponding to
LDA, HDA and VHDA\cite{ourVHDA,ourLDAHDA}.  Here, we simulate a system of
$N=216$ molecules \cite{footnoteSize},
 and average the results over 16 independent trajectories.  We use periodic
 boundary conditions. Long range forces are treated using the reaction field
 method. 

To study the phase diagram of glassy water we perform three different kinds of simulations: (i) isobaric cooling/heating, (ii) isochoric cooling/heating, and (iii) isothermal compression/decompression.

During cooling/heating at both constant $\rho$ or $P$, we change $T$ by
rescaling the velocities of the molecules.  At every time step $\delta
t=1$~fs, we increase the $T$ by $\delta T=q\delta t$, 
where $q$ is the cooling/heating rate and $q=\pm 3 \times 10^{10}$~K/s.
  These values for $q$ have been
used in previous simulations to study the effect of cooling and heating rates on the glass transition temperature $T_g=T_g(P)$ \cite{ourSHADOW,ourPRE-RC}.  The
glass obtained with a cooling rate $q=-3 \times 10^{10}$~K/s behaves, upon heating at $q=+3 \times 10^{10}$~K/s, as a slow-cooled glass as found in experiments showing no signs of hyperquenching effects. On the other hand, the lowest cooling/heating rates accessible nowadays in computer simulations  are $|q| \approx 10^{10}$~K/s
\cite{french}.
 
For the compression/decompression 
simulations at $T>0$~K we perform MD simulations at constant
 $\rho$ for intervals of $1$~ps. At the end of each interval of $1$~ps, we increase $\rho$ by 
$\delta \rho = \pm 5\times 10^{-5}$~g/cm$^3$ ,so our 
compression/decompression rate is $\delta \rho / \delta t= \pm 5\times
10^{-5}$~g/cm$^3$/ps.  This value was already used to study the potential
energy landscape for the LDA-HDA transformation\cite{ourLDAHDA}. 
 These $\rho$ changes are  performed by 
rescaling isotropically the coordinates of the molecular center of mass.
For the compression simulations at $T=0$~K, at each simulation step  we change $\rho$ by $\delta \rho = \pm 5\times 10^{-5}$~g/cm$^3$ and
 then minimize the energy. At each step $\rho$ is modified by  rescaling isotropically the center of mass of each molecule.

\section{LDA-HDA Transformation}
\label{LDAtoHDA}

We prepare LDA by hyperquenching at constant $\rho = 0.9$~g/cm$^3$ equilibrium liquid configurations obtained at $T=220$~K.  
We have chosen a glass obtained quenching a low $T$ liquid equilibrium configuration --- i.e. properly speaking a  HGW ----at $\rho=0.9$ g/cm$^3$ as starting LDA structure to guarantee that the chosen configuration is a low $\rho$ glass with an optimized tetrahedral network of hydrogen bonds.
To obtain HDA, we compress LDA at constant $T$ \cite{ourLDAHDA} from $\rho
= 0.9$~g/cm$^3$ to $\rho > 1.5$~g/cm$^3$.  The $T$ is fixed at
$T=0$, 77, 100, 120, 140, 160, and 170~K. These $T$ are below $T_g(P)$.
Evaluation of the specific heat upon heating the glass with a heating rate of
$q=+3 \times 10^{10}$~K/s to the liquid phase indicates that $T_g(P)
>170$~K for all $P$s studied.

\begin{figure}[p]
\narrowtext 
\centerline{
\hbox {
  \vspace*{0.5cm}  
  \epsfxsize=15cm
  \epsfbox{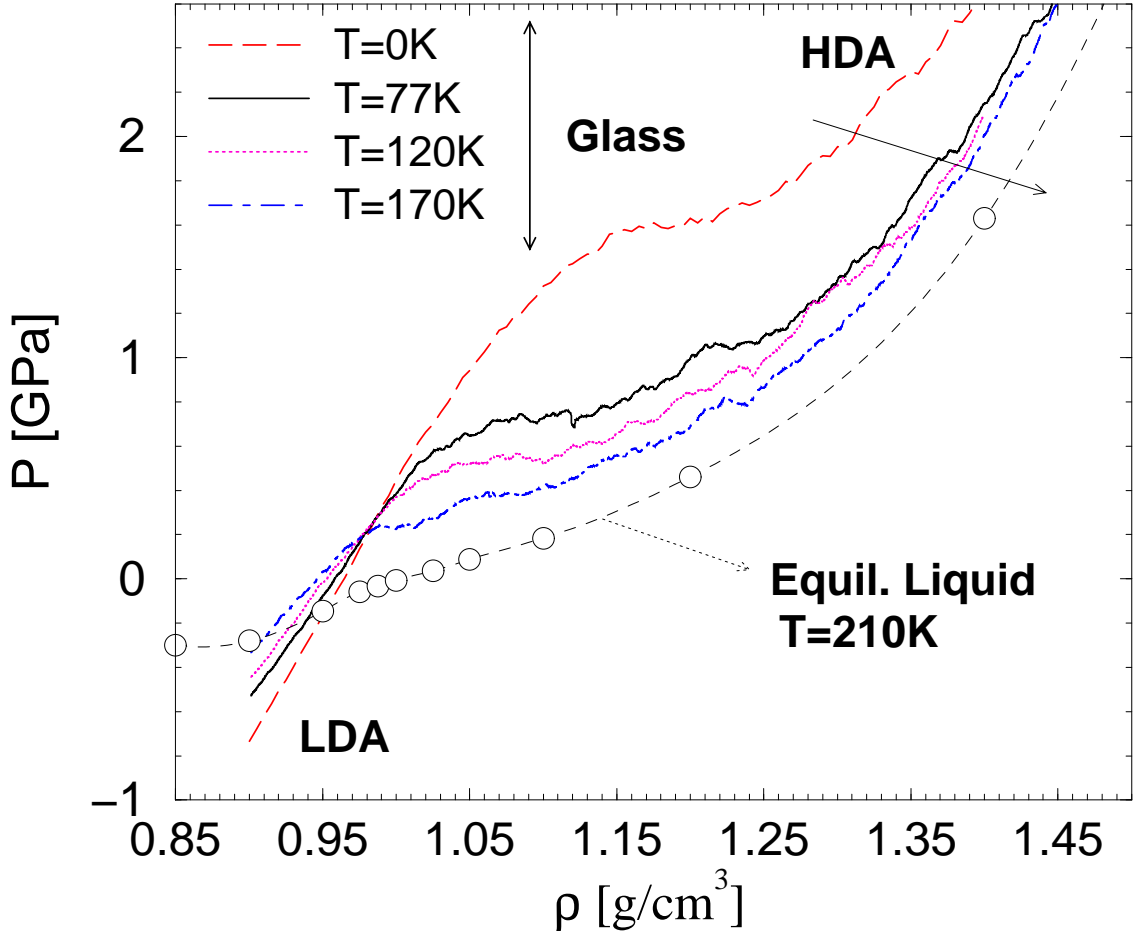}
}}
\vspace*{0.5cm}
\caption{Transformation of LDA to HDA obtained by isothermal compression
  at $T=0$, $77$, $120$, and $170$~K. LDA was obtained by hyperquenching
  at constant $\rho$ equilibrium liquid configurations at $T=220$~K and
 $\rho =  0.9$~g/cm$^3$.  The transformation starts approximately at ($\rho_c
  =1.03$~g/cm$^3$, $P_c=0.7$~GPa) at $T=0$~K and shift with $T$ to ($\rho_c
  =0.97$~g/cm$^3$, $P_c=0.13$~GPa) at $T=170$~K. We also
  show data for equilibrium liquid at
  $T=210$~K (from Ref.~{\protect \cite{poolepre}}), one of the lower $T$
  accessible for equilibrium liquid simulations. 
We note that glass isotherms cross at
 ($\rho \approx 0.98$~g/cm$^3$, $P \approx 0.2$~GPa) and that as
 $T$ increases, the glass isotherms approach the liquid isotherm. }
\label{ldahda}
\end{figure}

In Fig.~\ref{ldahda} we show $P(\rho)$ during isothermal compression at four
 of the simulated $T$. We also show equilibrium data taken from the liquid phase simulations reported in Ref.~\cite{pooleSPC/E}. From Fig.~\ref{ldahda} we note:

\begin{itemize}

\item[{(i)}] For all $T \leq 170$K isotherms show a transformation from LDA
  to  HDA. The linear increase of $P(\rho)$ crosses to a transition region 
around  $\rho_c \approx 1.03$~g/cm$^3$ and $P_c \approx 0.7$~GPa for 
 $T=0$~K and  $\rho_c \approx 0.97$~g/cm$^3$ 
and $P_c \approx 0.13$~GPa for $T=170 K$.
           The LDA-HDA
           transition starts at higher $\rho$ and $P$ on
           decreasing $T$. This feature has been observed in computer
           simulations of different models of water, including ST2 
           \cite{poole,poolepre} and TIP4P \cite{poolepre},
           and agrees with experiments \cite{mishimaJCP,whalley}.

\item[{(ii)}] All the compression curves at $T>0$~K collapse
           to a single curve at
           very high $\rho$, suggesting that there is a single HDA
           state at {\it very}  high $P$. 
            This is not the case for LDA because the $P$ of LDA 
           while compressing at $\rho=0.9$~g/cm$^3$ depends on $T$.
           Furthermore, a study of
           the potential energy landscape during the LDA-HDA
           transformation shows clear differences between two LDA samples
           at low $P$ and $T$~\cite{ourLDAHDA}.
 
\item[{(iii)}]  The isotherms in the glassy states cross each
 other at approximately $\rho =0.98$~g/cm$^3$ and $P=0.2$~GPa
 (see also Fig.~\ref{ldahdaLDA}). This could suggest
 (if these out of equilibrium isotherms could be interpreted as an effective
 equation of state of the glass) that $\rho$ increases
upon isobaric heating the glass at high $P$,
while $\rho$ decreases upon isobaric heating the glass at low $P$. 

\item[{(iv)}] 
We note that as the compression temperature increases from
           $T=0$~K to $T=170$~K the isotherms in the glassy state 
          approach the liquid isotherm for $T=210$~K.  Accordingly, we
           see that at low $\rho$ (LDA, e.g. $\rho =0.9$~g/cm$^3$)
           increasing $T$ increases $P$, while 
           at high $\rho$ (HDA, e.g. at $\rho=1.3$~g/cm$^3$) 
           increasing $T$ decreases $P$.

\end{itemize}

\section{Isobaric Heating of Glassy Water. Very High Density  Amorphous Ice and ``other'' Amorphous Ices.}
\label{HDAtoVHDA} 

\subsection{Heating at High $P$}
\label{secA}

Experimentally, VHDA is obtained by isobaric
heating of HDA ($T=77$~K, $\rho=1.33$~g/cm$^3$) at $P=1.1$~GPa and HDA
($T=77$~K, $\rho=1.45$~g/cm$^3$) at $P=1.9$~GPa \cite{loerting}.
 The products are VHDA ($T\approx 160$~K, $\rho\approx
1.37$~g/cm$^3$) and VHDA ($T\approx 177$~K, $\rho\approx 1.51$~g/cm$^3$),
respectively. Simulations with the SPC/E model also show a transformation between HDA to another glassy state identified with VHDA
\cite{ourVHDA}.  These simulations also suggest that HDA and VHDA are not
different glasses, but rather that VHDA is the result of the relaxation of
HDA upon heating. Furthermore, it is found that VHDA is the
same glass obtained by cooling HDL \cite{french,ourVHDA}.

\begin{figure}[p]
\narrowtext 
\centerline{
\hbox {
  \vspace*{0.5cm}  
  \epsfxsize=15cm
  \epsfbox{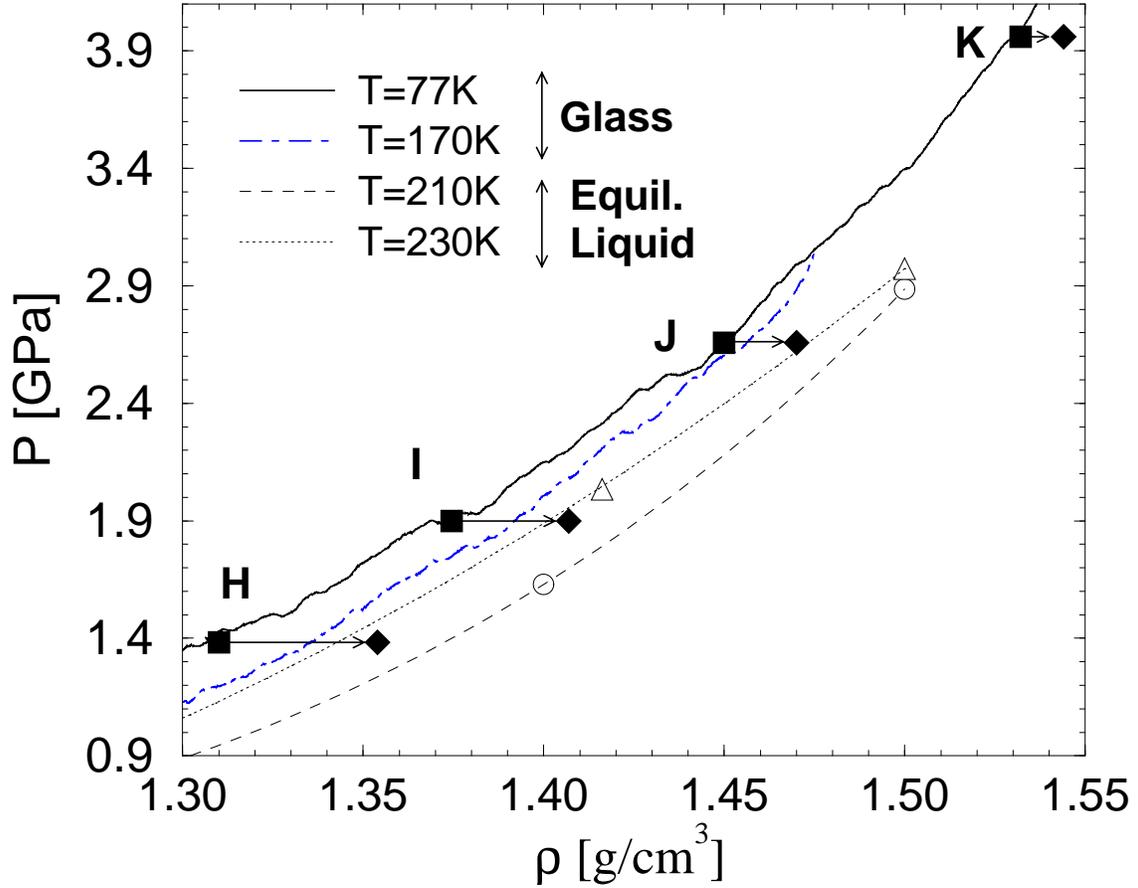}
}}

\vspace*{0.5cm}
\caption{Magnification of Fig.~\ref{ldahda} corresponding to densities
  of HDA. To test the presence of VHDA we heat at constant $P$
  configurations corresponding to state points H, I, J and K in the figure (see also
  Table~\ref{points}). Upon heating from $T=77$~K (squares) up to $T\approx
  170$~K (diamonds), the four state points move to higher densities crossing
 the glass isotherm for $T=170$~K and approaching the $T=210$~K liquid
 isotherm (see arrows).
}
\label{ldahdaHDA}
\end{figure}

\begin{figure}[p]
\narrowtext 

\centerline{
\hbox {
  \vspace*{0.5cm}  
  \epsfxsize=9cm
  \epsfbox{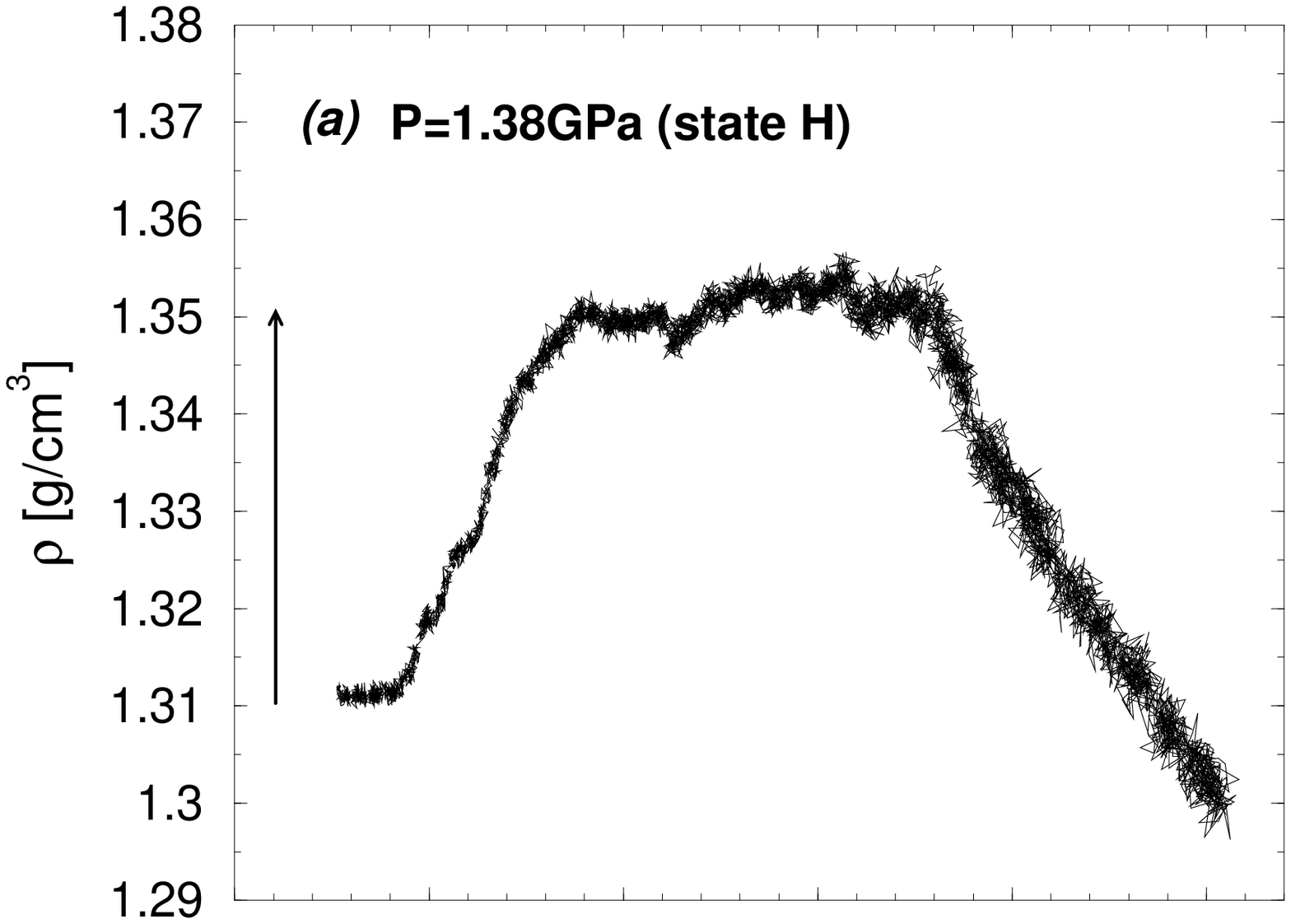}
  \hspace*{0.1cm}
  \epsfxsize=9cm
  \epsfbox{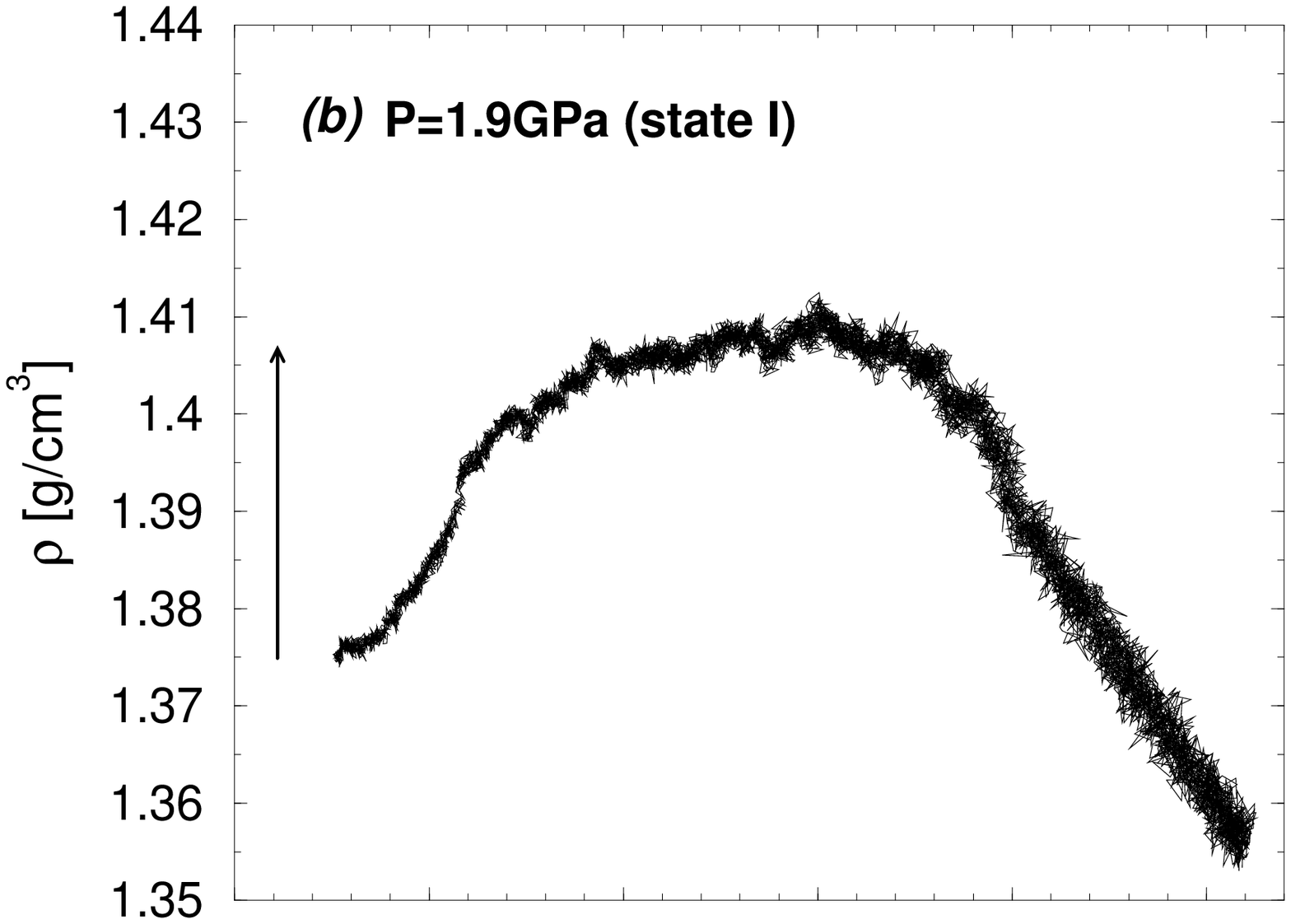}
}}
\centerline{
\hbox {
  \vspace*{0.5cm}  
  \epsfxsize=9cm
  \epsfbox{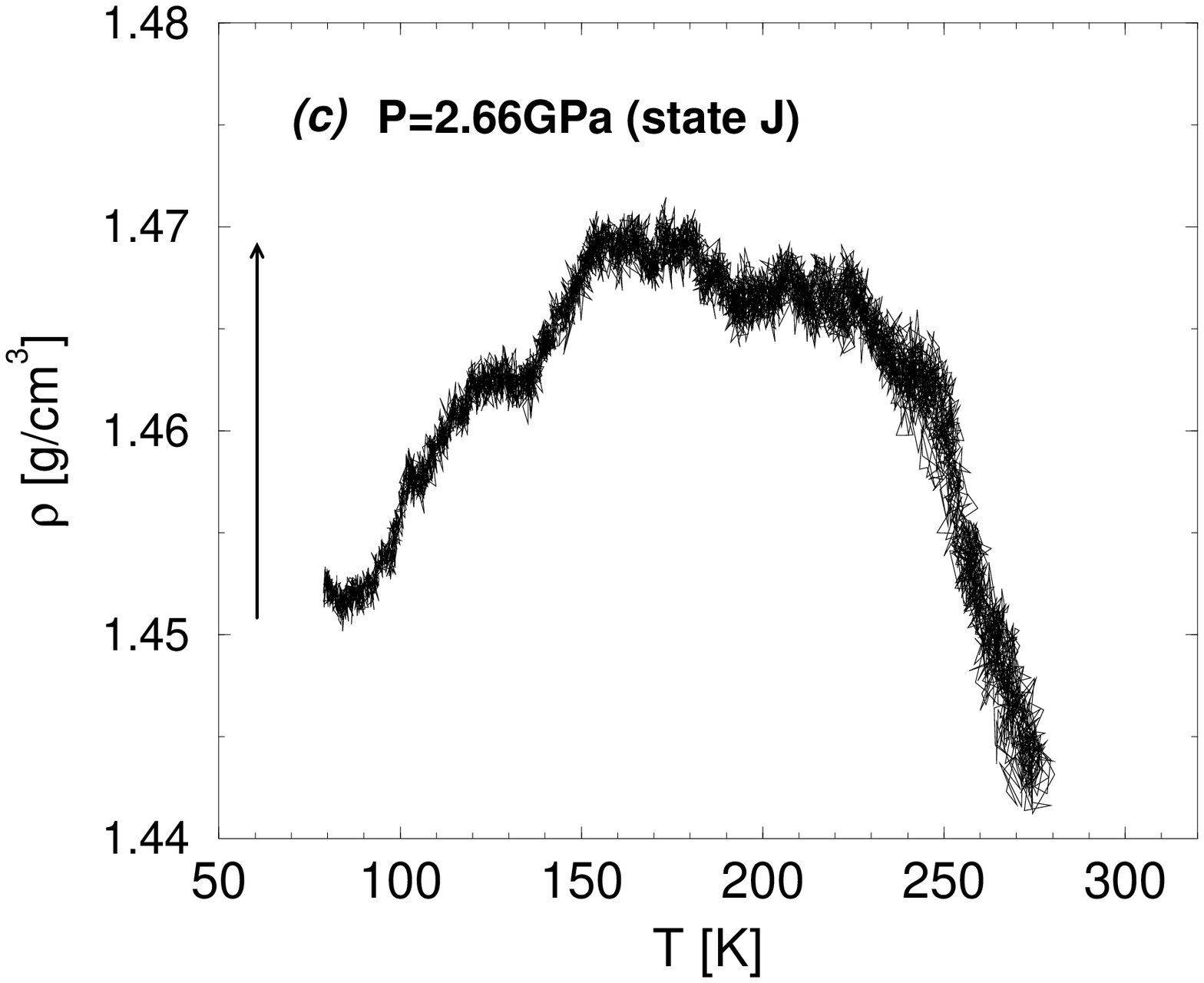}
  \hspace*{0.1cm}
  \epsfxsize=9cm
  \epsfbox{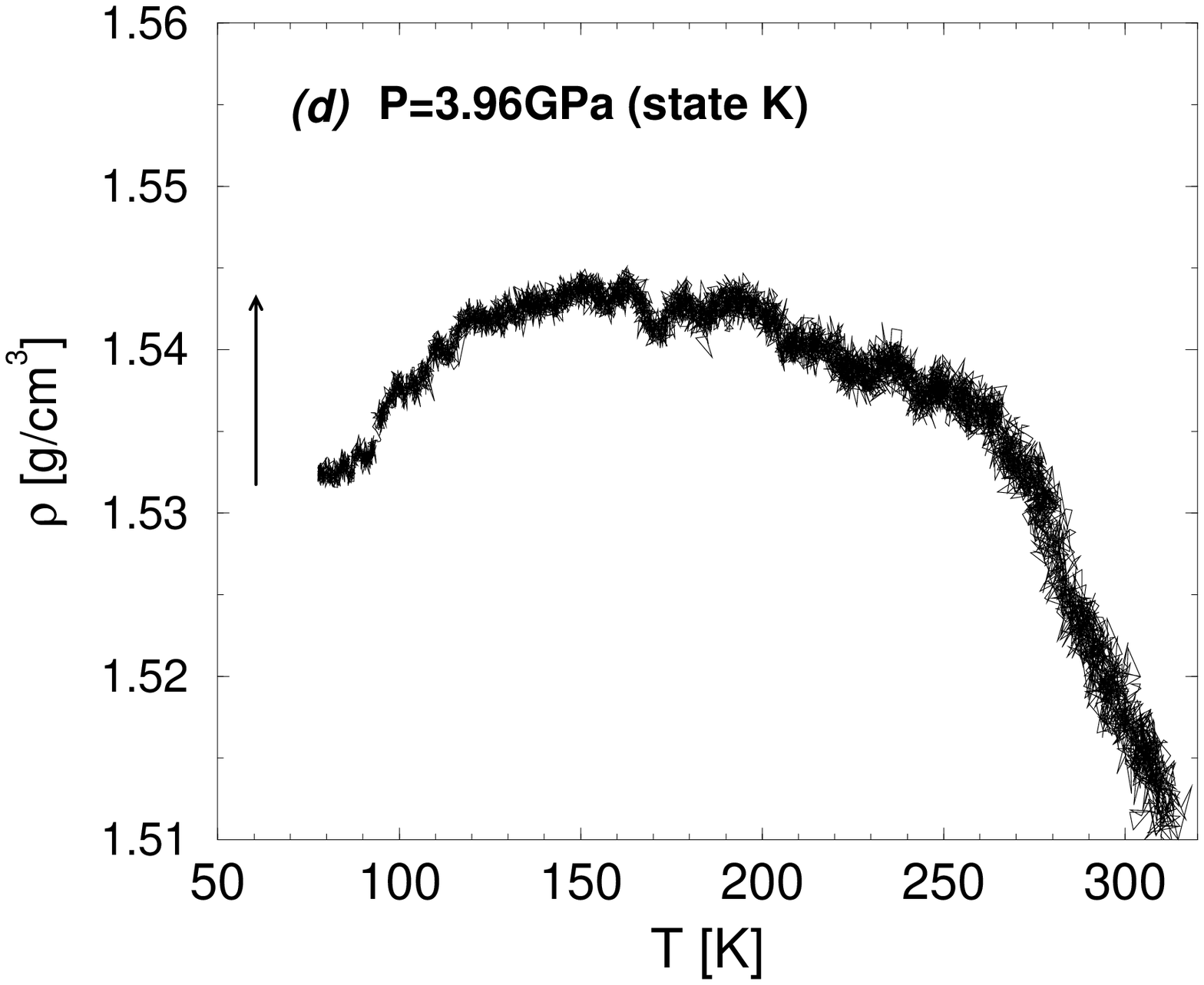}
}}
\vspace*{0.5cm}
\caption{Evolution of $\rho$ upon isobaric heating for the four state points
  H, I, J, K corresponding to HDA, indicated in Fig.~\ref{ldahdaHDA}. In
  all cases, $\rho$ increases with $T$ before the liquid phase is
  reached.  The denser glass to which HDA transforms is identified as
  the VHDA obtained in experiments. 
Arrows indicate the increase in $\rho$ when heating the glass from $T=77$~K
  up to $T \approx 170$~K. The increase in $\rho$ is smaller
  as $P$ increases.}
\label{heat1.3X}
\end{figure}

We obtain HDA by compressing LDA at $T=77$~K as explained in the last section.  We confirm the transformation of HDA to a denser glass by 
isobaric heating of HDA at different $P$s.   The HDA configurations correspond to four different states, indicated by state points H, I, J and K in Fig.~\ref{ldahdaHDA} (see Table~\ref{points}). 
For the four cases studied the densities upon heating are shown in Fig.~\ref{heat1.3X}. In all cases, $\rho$ increases when heating
the glass from $T=77$~K to $T\approx 170$~K still below $T_g(P)$. On
increasing $T$, the glass melts
to the equilibrium liquid. The $\rho$ increase is smaller the higher the $P$ and it is indicated by arrows in Fig.~\ref{ldahdaHDA}. Note
that the increase in $\rho$ of $\approx 0.04$~g/cm$^3$ that we find at
$P=1.38$~GPa is very close to the estimated experimental values\cite{loerting} of $0.04$~g/cm$^3$ and $0.06$~g/cm$^3$ for
isobaric heating at $P=1.1$ and $P=1.9$~GPa, respectively.

The $\rho(T)$ values  shown in Fig.~\ref{heat1.3X} at high
$T$   are the ones obtained in the equilibrium liquid.  
Indeed, at $P=1.38$~GPa  and  $T=230$~K the relaxation time $\tau$ is $\tau
 \approx 100$~ps \cite{francislong2}.
 Heating at a rate of $q_h=3 \times 10^{10}$~K/s, the $T$-change during
$100$~ps is about 3~K. Therefore, the system has time to relax before $T$ changes considerably and the $\rho$ measured is the same as in
the equilibrium liquid. The decrease of $\rho$ observed in Fig.~\ref{heat1.3X} for $T \geq 230$~K (in the liquid phase) is in agreement with the location of the   isotherms for $T=210$~K and $T=230$~K at high $P$ shown in Fig.~\ref{ldahdaHDA}.

\begin{figure}[p]
\narrowtext 

\centerline{
\hbox {
  \vspace*{0.5cm}  
  \epsfxsize=12cm
  \epsfbox{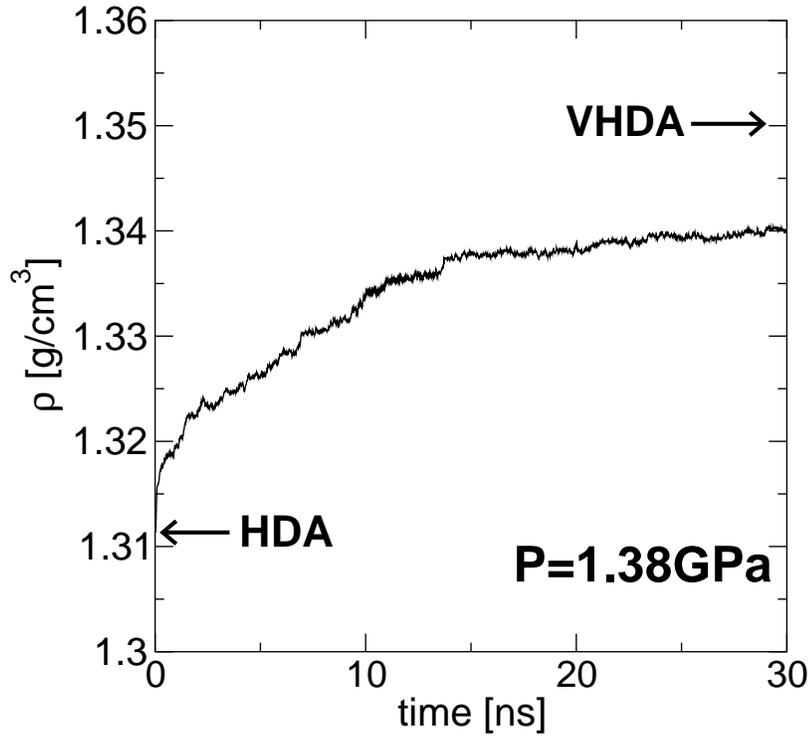}
}}

\vspace*{0.5cm}
\caption{Effect of aging configurations of HDA corresponding to state point H in
Fig.~\ref{ldahdaHDA}. With time, the $\rho$ increases approaching the
$\rho$ corresponding to VHDA at $P=1.38$~GPa. 
This finding is consistent with  the view of VHDA as the result of relaxation
 of HDA upon isobaric heating.}
\label{rho-t}
\end{figure}

Computer simulations show that the final glass connected to  the liquid phase
 by cooling at high $P$ is  VHDA \cite{ourVHDA}. We argue \cite{ourVHDA} that HDA is less
 stable than VHDA and it is a kinetically
trapped glass that could not relax during the compression of LDA.
The hypothesis that HDA is a kinetically
trapped glass was already proposed based on recent computer simulations
performed with the TIP4P model \cite{parrinello} and it was first 
 proposed by Mishima who observed that HDA samples
(made by different $P$-$T$ conditions) heated or annealed to $130-150$~K at
$1-1.5$~GPa were characterized by identical X-ray patterns \cite{mishima96}.
 To provide further support to this hypothesis, we age configurations corresponding to 
state point H in Fig.~\ref{ldahdaHDA} before heating, i.e. at $P=1.38$~GPa and
$T=77$~K. The evolution of $\rho$ with time $t$ is shown in
 Fig.~\ref{rho-t}. As $t$  goes on, $\rho$ increases  approaching the value
 of $\rho$ corresponding to VHDA at $P=1.38$~GPa.
Another support to the hypothesis that  VHDA can be interpreted as  `relaxed
HDA' comes from the study of  the effect of the compression $T$ in the
LDA$\rightarrow$HDA transformation. If HDA is less stable than VHDA, we
should find that (for a given final $P$ after compression 
of LDA) the larger the  compression $T$, the closer is HDA to the
corresponding VHDA. In other words, the higher the compression $T$ of LDA,
 the more the system can relax to a more stable state.  Accordingly,
 Fig.~\ref{ldahdaHDA} shows that as the compression $T$ is increased, the 
 compression  isotherms starting from LDA 
shift (for a given $P$)  to higher $\rho$, approaching VHDA.

\subsection{Heating at Intermediate $P$}

\begin{figure}[p]
\narrowtext 

\centerline{
\hbox {
  \vspace*{0.5cm}  
  \epsfxsize=15cm
  \epsfbox{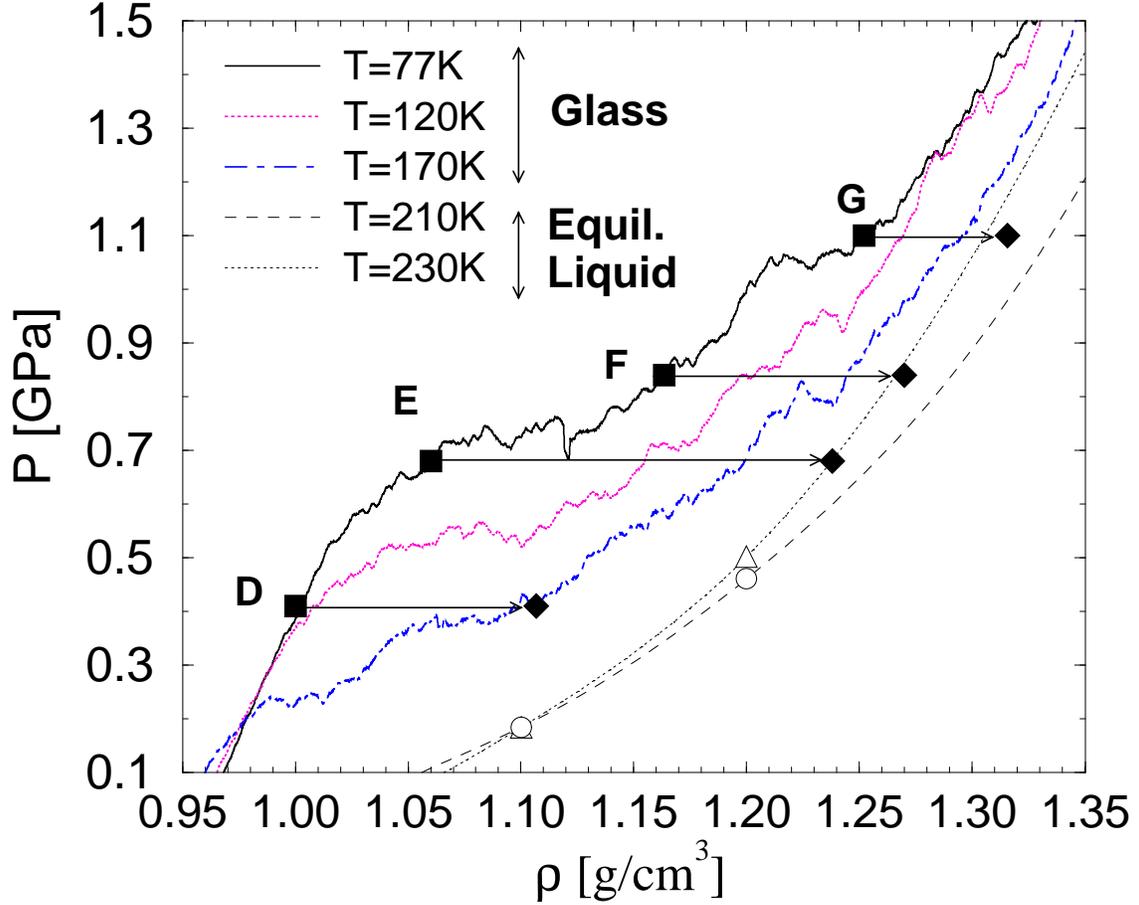}
}}

\vspace*{0.5cm}
\caption{Magnification of Fig.~\ref{ldahda} for intermediate densities
  in the LDA-HDA transition. We heat isobarically glass configurations
  corresponding to amorphous ices obtained during the transformation of
 LDA to HDA (state points D, E, F), and one
  configuration corresponding to LDA (state point G; see also
  Table~\ref{points}). All state points shift to higher $\rho$
  upon heating (see arrows). Squares indicate the $\rho$ of the 
starting glasses at $T=77$~K while 
diamonds indicate the final $\rho$ reached at $T\approx 170$~K 
  before the glass transition. We also show the $P-\rho$ values
  corresponding to two liquid isotherms (from Ref.~{\protect
  \cite{poolepre}}).   }
\label{ldahdaHDALDA}
\end{figure}

\begin{figure}[p]
\narrowtext 

\centerline{
\hbox {
  \vspace*{0.5cm}  
  \epsfxsize=9cm
  \epsfbox{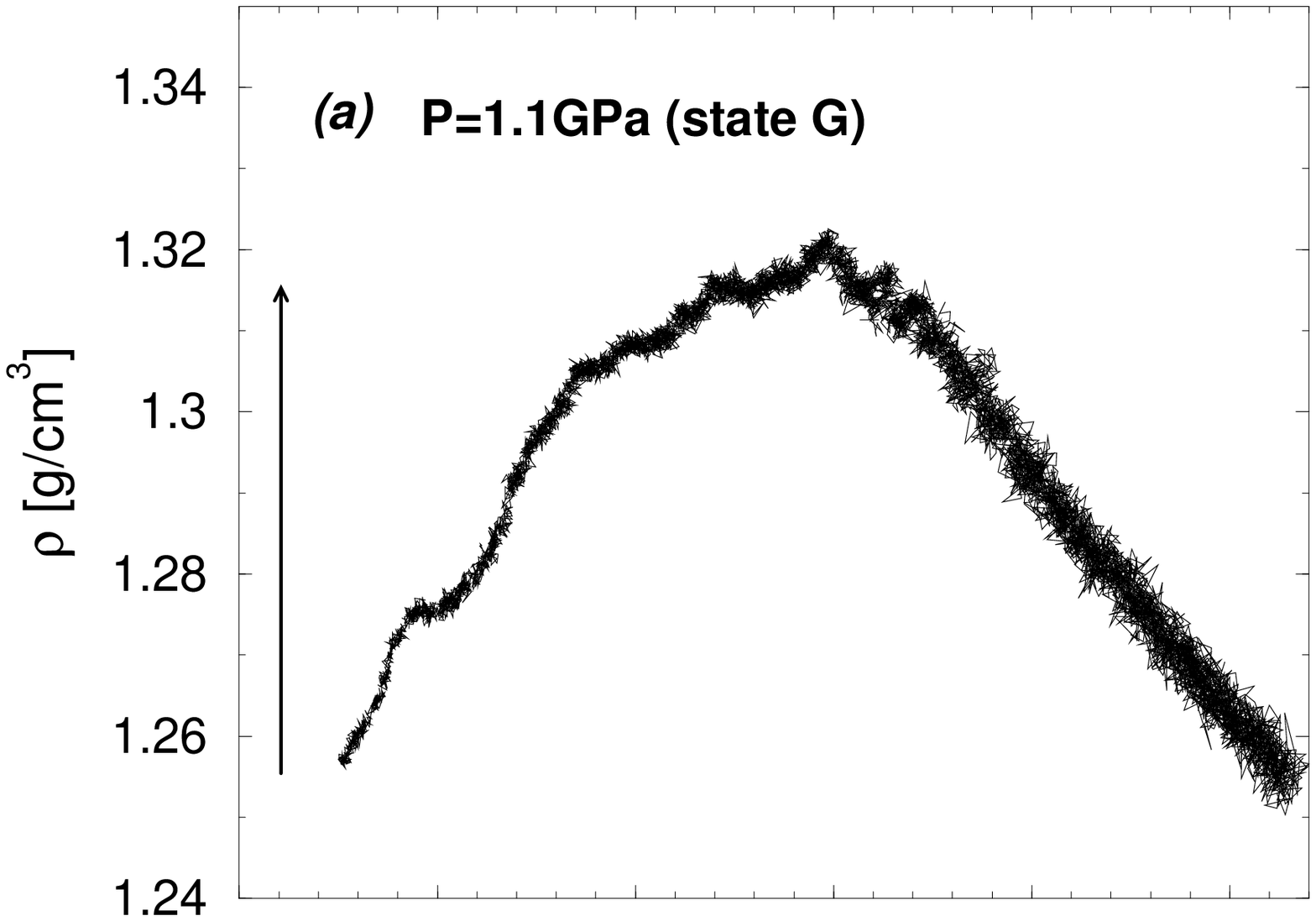}
  \hspace*{0.1cm}
  \epsfxsize=9cm
  \epsfbox{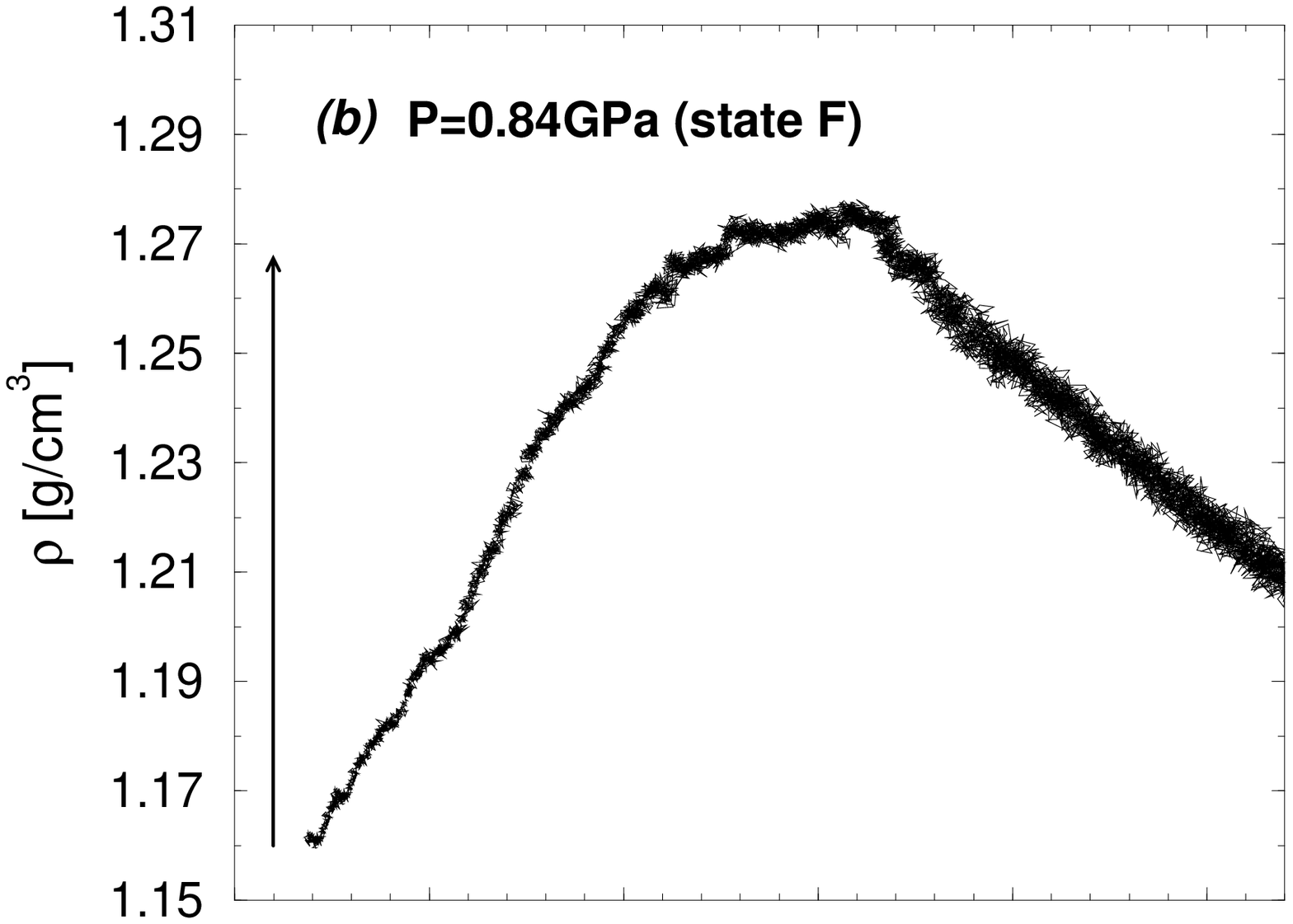}
}}
\centerline{
\hbox {
  \vspace*{0.5cm}  
  \epsfxsize=9cm
  \epsfbox{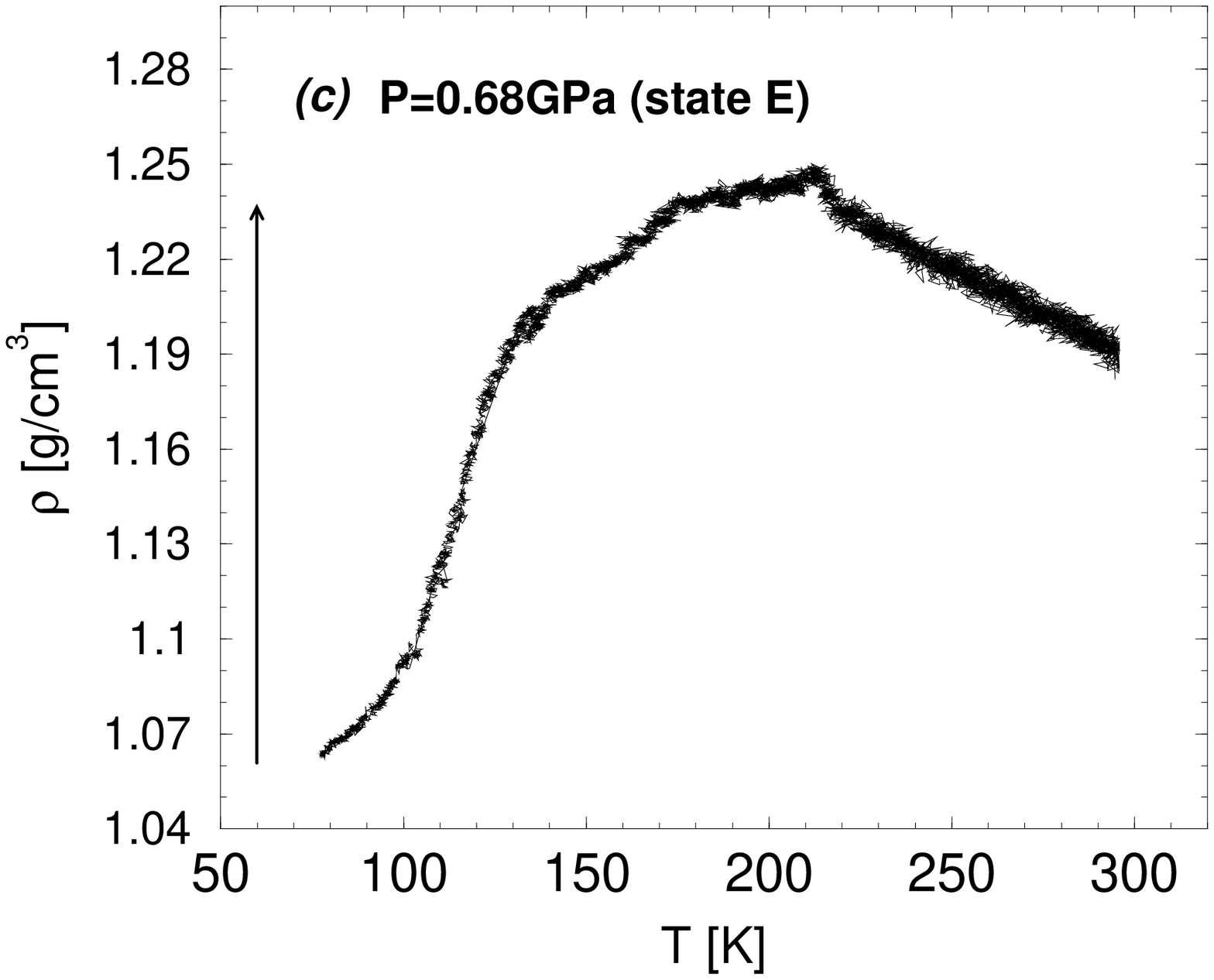}
  \hspace*{0.1cm}
  \epsfxsize=9cm
  \epsfbox{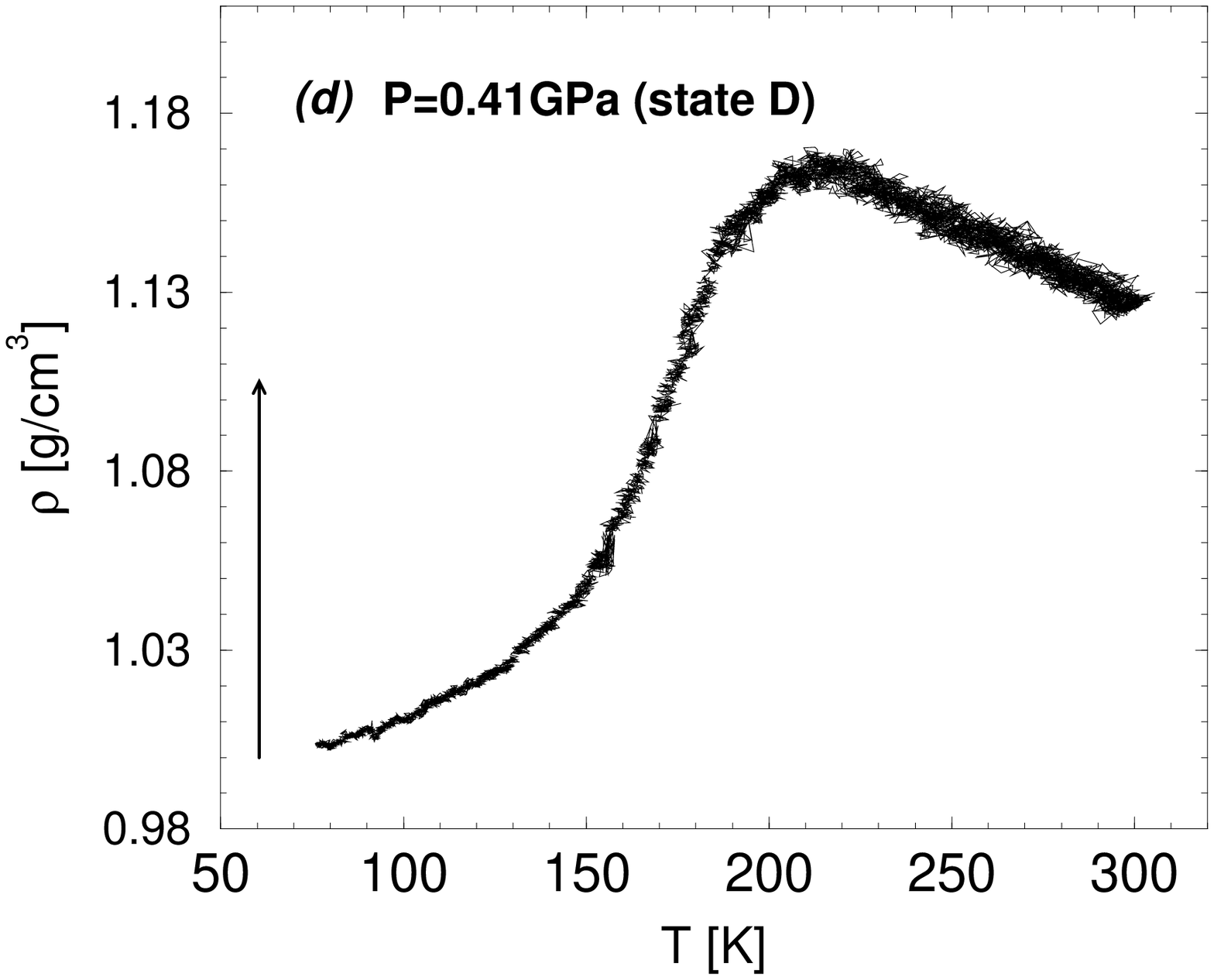}
}}
\vspace*{0.5cm}
\caption{Evolution of $\rho$ upon isobaric heating for the four state points
  D, E, F, and G indicated in Fig.~\ref{ldahdaHDALDA}. 
Arrows indicate the increase in $\rho$ when heating the glass from $T=77$~K
  up to $T \approx 170$~K. The smallest
  increase in $\rho$ occurs for the heating of configurations indicated
  by state point G and it is larger for heatings at lower $P$s.}
\label{heat1.00-1.25}
\end{figure}

We explore now the effect of isobaric heating of the glass at intermediate $P$, in the LDA-HDA transition region, indicated by state points D, E, F, and G in Fig.~\ref{ldahdaHDALDA} (see Table~\ref{points}). The lower the $P$ of these HDA forms, the more contaminated they are by LDA.  Upon
isobaric heating from $T=77$~K up to $T \approx 170$~K, the states indicated by state points D, E, F, and G shift to higher densities as indicated by the arrows in
Fig.~\ref{ldahdaHDALDA}. The evolution of the $\rho$ upon heating these four glasses are shown in Fig.~\ref{heat1.00-1.25}.  The situation is
analogous to the HDA-VHDA transformation discussed above. The increase
in $T$ allows these glasses to relax to more stable glass configurations
characterized by a higher $\rho$. The final glasses found after heating
at these intermediate $P$ correspond to the ``VHDA'' obtained at
high $P$s.  We note also that at a given intermediate $P$
(e.g. $P=0.8$~GPa), the isotherms obtained in the LDA-HDA
transformations shift to higher $\rho$ as the compression $T$ increases. This
is because, as discussed in Sec.~\ref{secA}, the higher the
compression $T$, the more the glass relaxes during compression.

\subsection{Heating at Low and Negative $P$}

\begin{figure}[p]
\narrowtext 

\centerline{
\hbox {
  \vspace*{0.5cm}  
  \epsfxsize=15cm
  \epsfbox{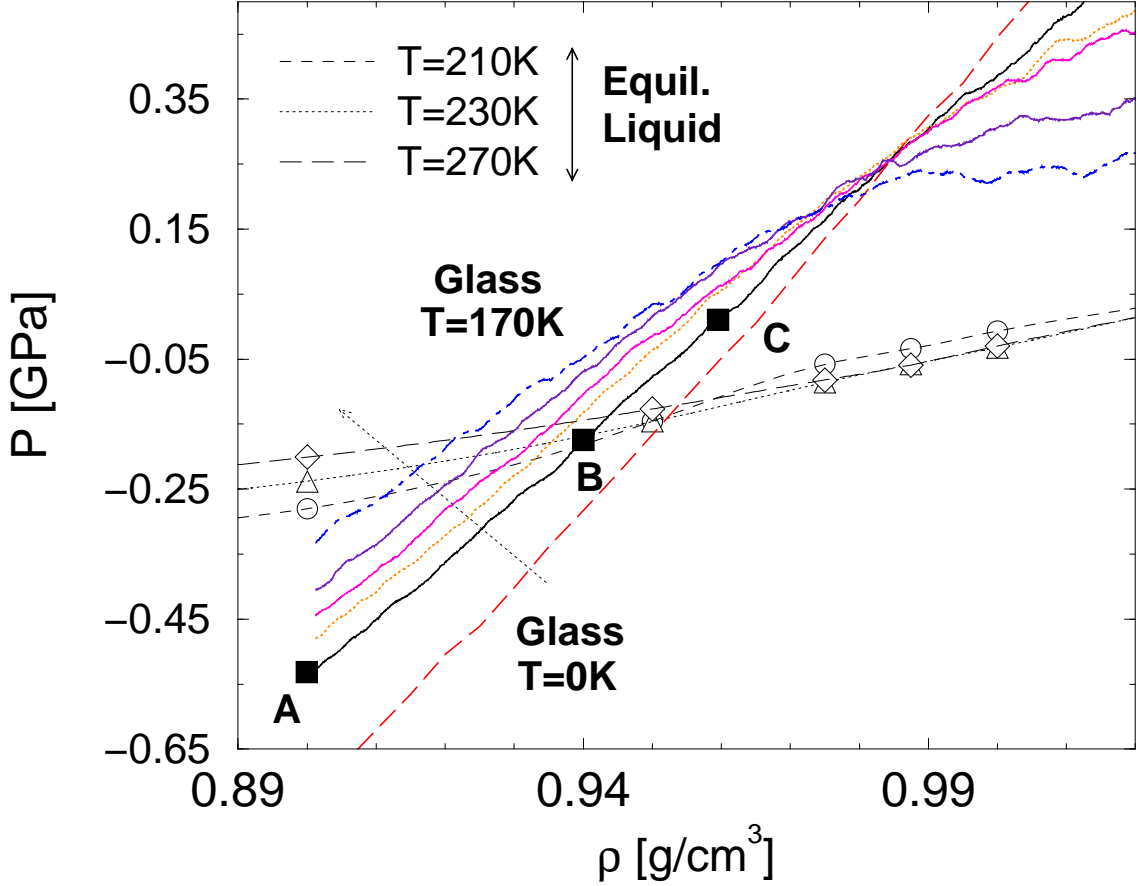}
}}

\vspace*{0.5cm}
\caption{Magnification of Fig.~\ref{ldahda} corresponding to densities
  of LDA. We heat at constant $P$ configurations corresponding to
  state points A, B, and C in the figure (see also Table~\ref{points}). We
  show all the glass isotherms simulated corresponding to $T=0$, $77$,
  $100$, $120$, $140$, and $170$K (see dashed arrow), and the liquid isotherms
  for $T=210$, $230$, and $270$~K (from Ref.~{\protect
  \cite{poolepre}}). 
 Note that glass isotherms cross each other 
at ($\rho \approx 0.98$~g/cm$^3$, $P \approx 0.2$~GPa).
  After heating and before the
  liquid phase is reached, state points  A, B, and C shift to lower densities.
 In the three cases, the
  state points finally approach the liquid isotherm at high $T$.}
\label{ldahdaLDA}
\end{figure}

\begin{figure}[p]
\narrowtext 

\centerline{
\hbox {
  \vspace*{0.5cm}  
  \epsfxsize=8.5cm
  \epsfbox{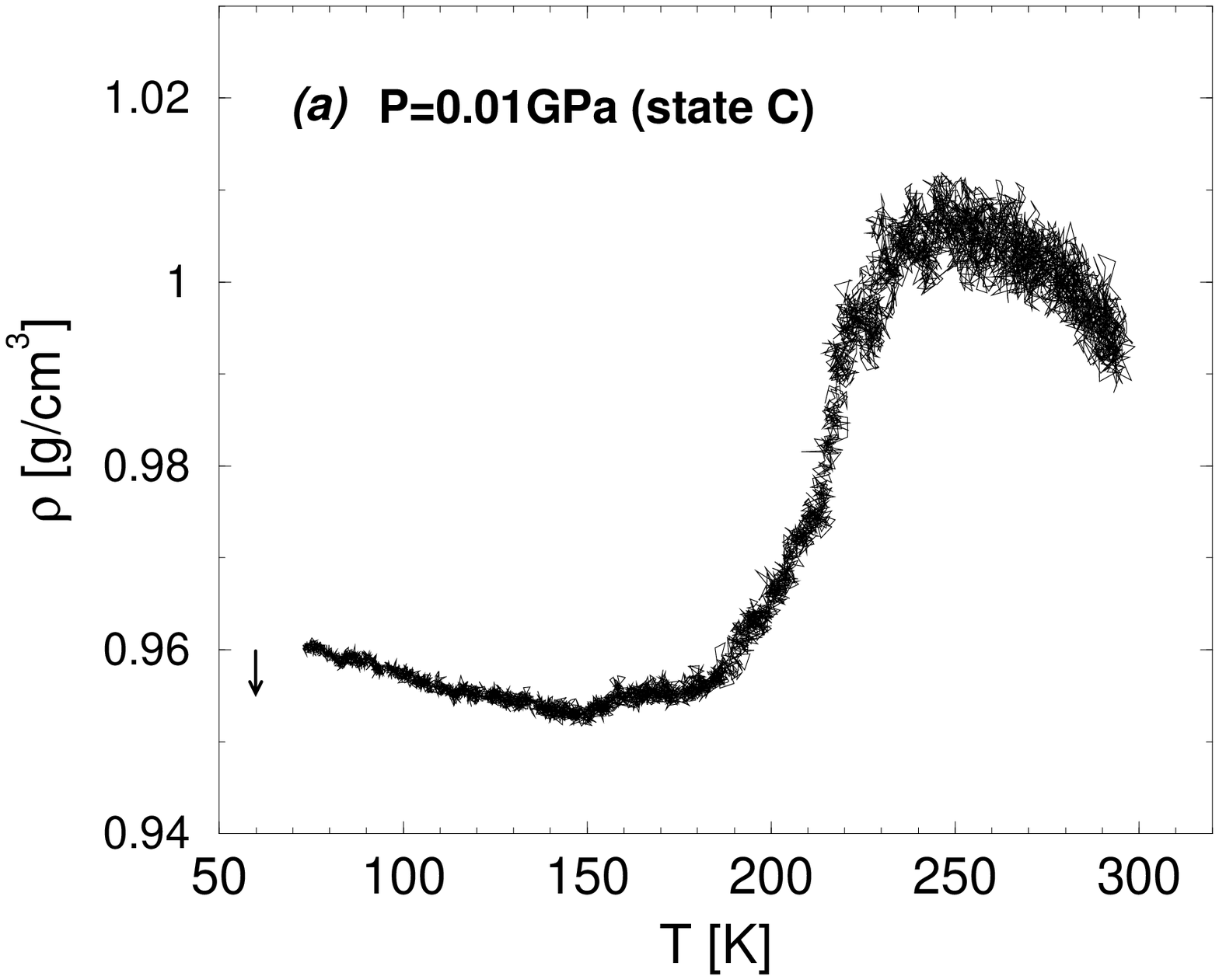}
  \hspace*{0.1cm}
  \epsfxsize=8.5cm
  \epsfbox{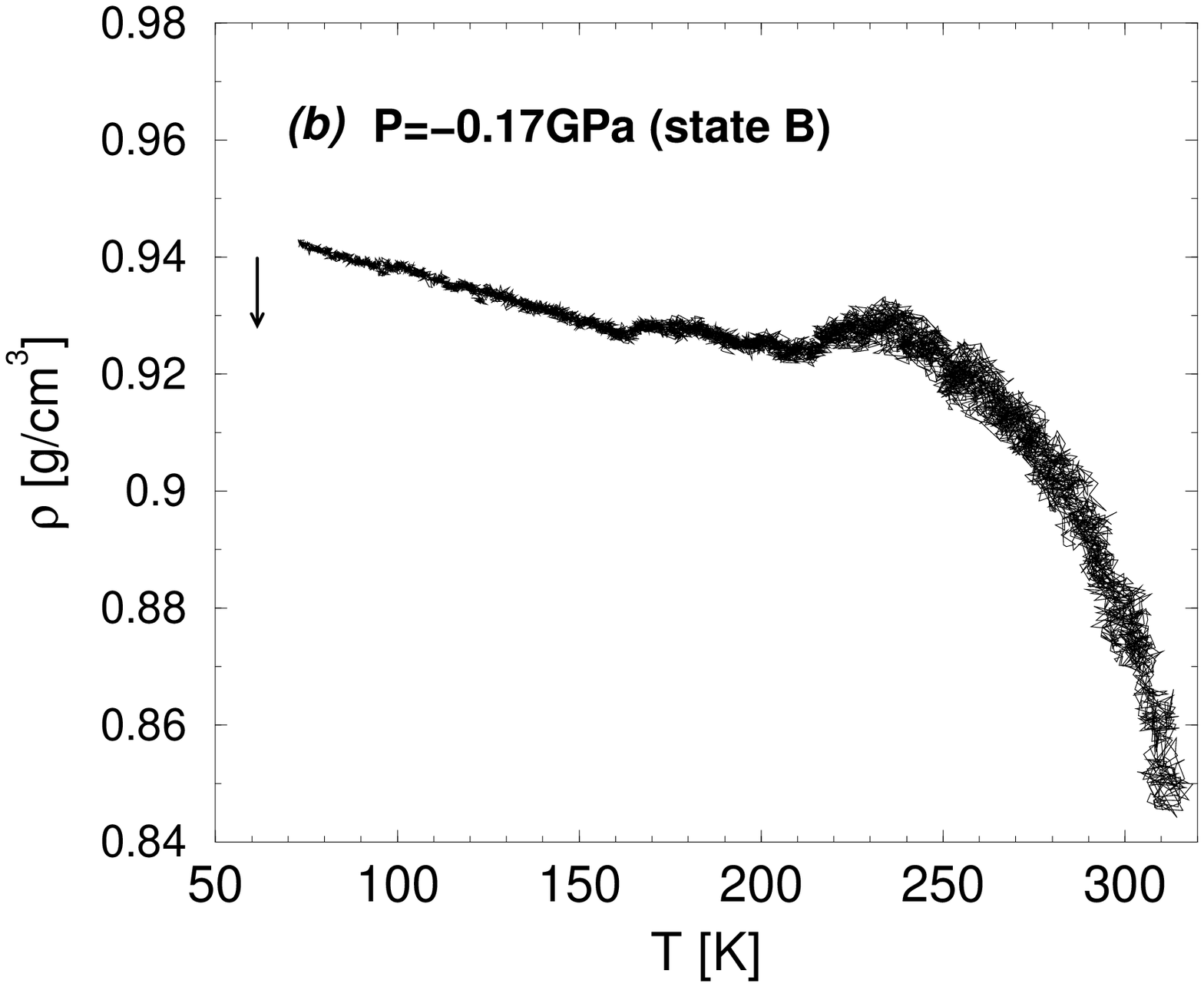}
}}
\centerline{
\hbox {
  \epsfxsize=8.5cm
  \epsfbox{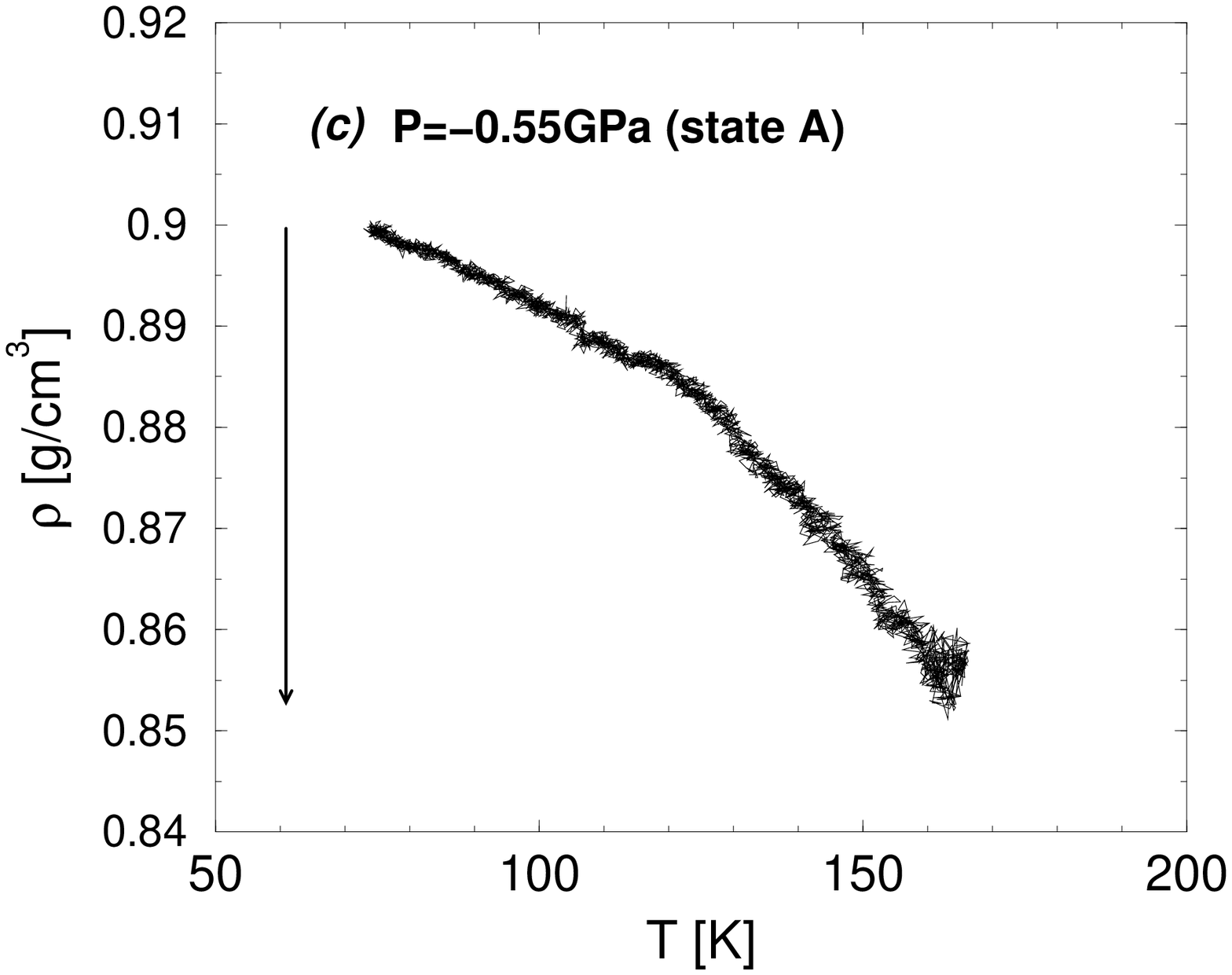}
}}

\vspace*{0.5cm}

\caption{Evolution of $\rho$ upon isobaric heating for the three state points
  A, B, and C indicated in Fig.~\ref{ldahdaLDA}. (a) Upon heating at
  $P=0.01$~GPa, $\rho$ decreases from $0.96$~g/cm$^3$ at $77$~K to
  $\rho \approx 0.952$~g/cm$^3$ at $T=150$~K (see arrow) 
and state point C moves through
  the glass isotherm in Fig.~\ref{ldahdaLDA}. Then $\rho$ increases and
  state point C shifts back to higher densities approaching the $T=230$~K
  equilibrium liquid isotherm.  (b) Upon heating the glass up to 
 $T \approx 170$~K at
  $P=-0.17$~GPa, $\rho$ decreases down to $\approx 0.93$~g/cm$^3$ (see arrow).
Upon farther heating up to $230$~K,
 $\rho$ does not change much probably
 because the $\rho$ of the equilibrium liquid at $P=-0.17$~GPa and $230$~K
is also $\approx 0.93$~g/cm$^3$.
(c) In this case, $\rho$
  decreases monotonically and state point A in Fig.~\ref{ldahdaLDA} shifts to
  lower densities.  The $\rho$ reaches very low values and the glass
  transforms into gas. Arrow indicates the increase in $\rho$ when heating
  the glass from $T=77$~K up to $T \approx 160$~K. }
\label{heat0.9-0.96}
\end{figure}

Isobaric heating at $P$s below the $P$ at which the glass isotherms cross
 ($P \approx 0.2$~GPa, $\rho \approx 0.98$~g/cm$^3$) presents some
 differences from the cases studied above.  We
discuss the heating from three LDA configurations indicated by state points
A, B and C in Fig.~\ref{ldahdaLDA} (see Table~\ref{points}). The heating
of LDA represented by state point C at $P=0.01$~GPa is shown in
Fig.~\ref{heat0.9-0.96}(a). We observe a decrease in $\rho$ from
$0.96$~g/cm$^3$ at $77$~K to $\rho \approx 0.952$~g/cm$^3$ at $T=150$~K,
and then an increase in $\rho$ until $T=230$~K (already in the liquid
phase).  This behavior of LDA upon heating at $P=0.01$~GPa is consistent with
the location of the isotherms in the glassy state. Isotherms in
 the glassy state shift to lower densities as $T$ increases,
 suggesting that the more
stable glasses at $P=0.01$~GPa are at lower densities.  Therefore, the
glass represented by state point C moves upon heating first to lower $\rho$
approaching the glass isotherm at $T=170$~K in the glassy state, and then
 shifts back to higher $\rho$ approaching the $T=230$~K equilibrium
liquid isotherm.  
 The relaxed glass obtained before the glass transition
is a ``very-low density glass'' (``VLDA'') and is conceptually analogous to the VHDA obtained at high $P$.
A more precise nomenclature for these glasses might be relaxed-LDA (RLDA) and
 relaxed-HDA (RHDA), respectively (see also \cite{urquidi2}).

A situation similar to the one discussed previously for state point C in
 Fig.~\ref{ldahdaLDA}  holds for state point B in Fig.~\ref{ldahdaLDA}, i.e. for
 the isobaric heating at $P=-0.17$~GPa of LDA ($\rho = 0.94$~g/cm$^3$).
The $\rho$ upon heating for this case is shown in Fig.~\ref{heat0.9-0.96}(b).
 We observe an initial decrease in $\rho$ down to $\approx 0.93$~g/cm$^3$ at
 $T \approx 170$~K approaching the value of $\rho \approx 0.92$~g/cm$^3$ corresponding to
 the glass isotherm at $T=170$~K. 
As $T$ increases up to $230$~K, a non monotonic 
density dependence arise, 
probably because the $\rho$ of the equilibrium liquid at $P=-0.17$~GPa and
 $230$~K is also $\approx 0.93$~g/cm$^3$. The final decrease in $\rho$ for
 higher $T$ is due to the fact that the liquid isotherms at $P=-0.17$~GPa in
 Fig.~\ref{ldahdaLDA} shift to lower $\rho$ as $T$ increases.

Finally, Fig.~\ref{heat0.9-0.96}(c) shows the effect of heating LDA at
 $P=-0.55$~GPa, indicated by state point A in Fig.~\ref{ldahdaLDA}.  In this
case, $\rho$ decreases abruptly upon heating and LDA transforms into gas
 (i.e., it ``sublimates'') because there are no liquid isotherms at this $P$.

\section{Recovering of HDA and VHDA at ambient $P$. A continuum of  glasses.}
\label{recoveredGLASSES}

In this section, we investigate the effects of isothermal expansion of 
 HDA and VHDA at low $T$ down to $\rho =0.85$~g/cm$^3$.
  The structure of HDA and VHDA has been described at $T=77$~K and $P=0.01$~GPa\cite{loerting,finneyLDAHDA}. To compare our results with the experiments, we also recover HDA and VHDA at these conditions.  For
both HDA and VHDA, the system is decompressed at constant $T=77$~K
(in the case of VHDA, we first bring the system back to $T=77$~K by keeping
 the $P$ constant).
 We recover the configurations of HDA represented by state points H, I, J, and K in Fig.~\ref{ldahdaHDA}, and the corresponding configurations of VHDA.

\begin{figure}[p]
\narrowtext 

\centerline{
\hbox {
  \vspace*{0.5cm}  
  \epsfxsize=9cm
  \epsfbox{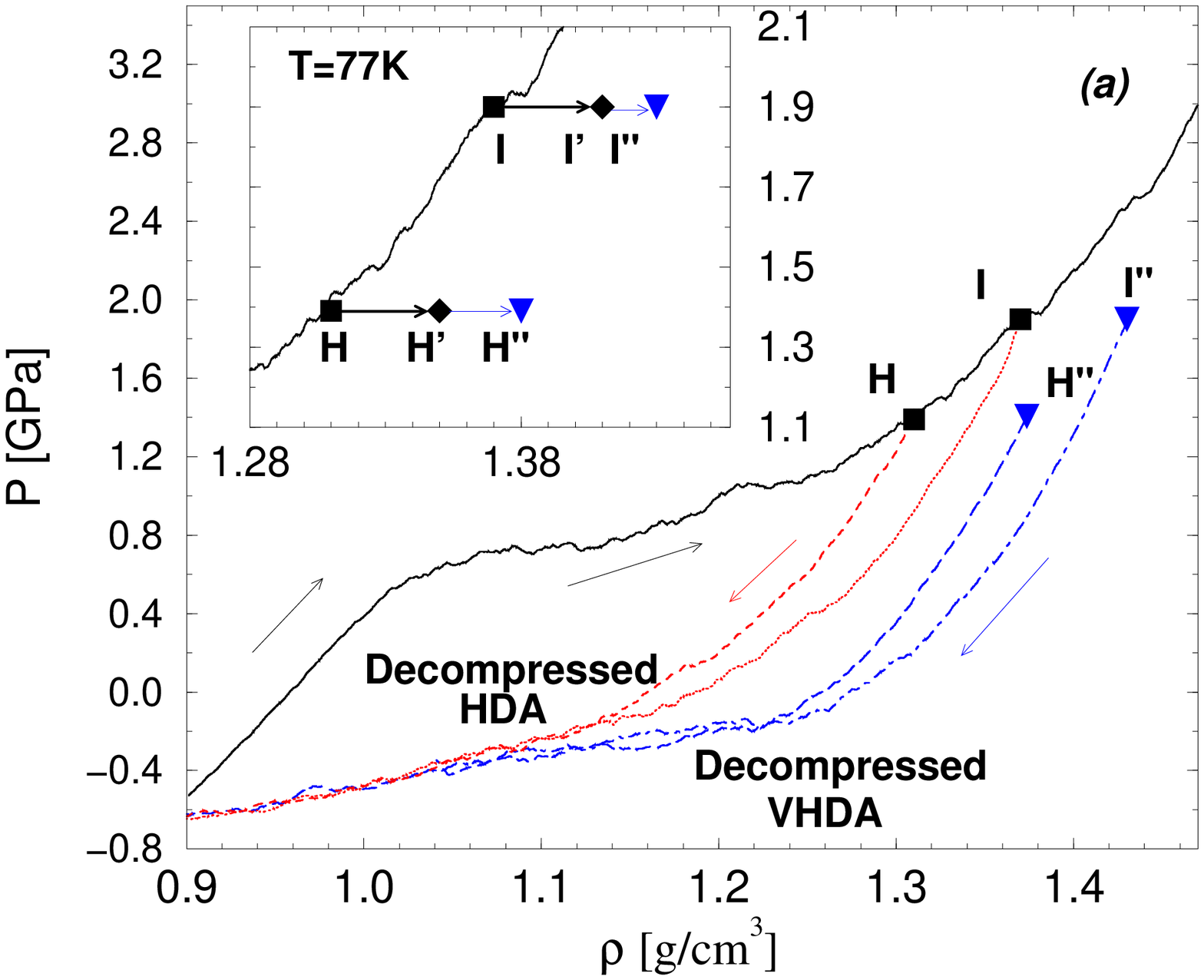}
}}
\centerline{
\hbox {
  \vspace*{0.5cm}  
  \epsfxsize=9cm
  \epsfbox{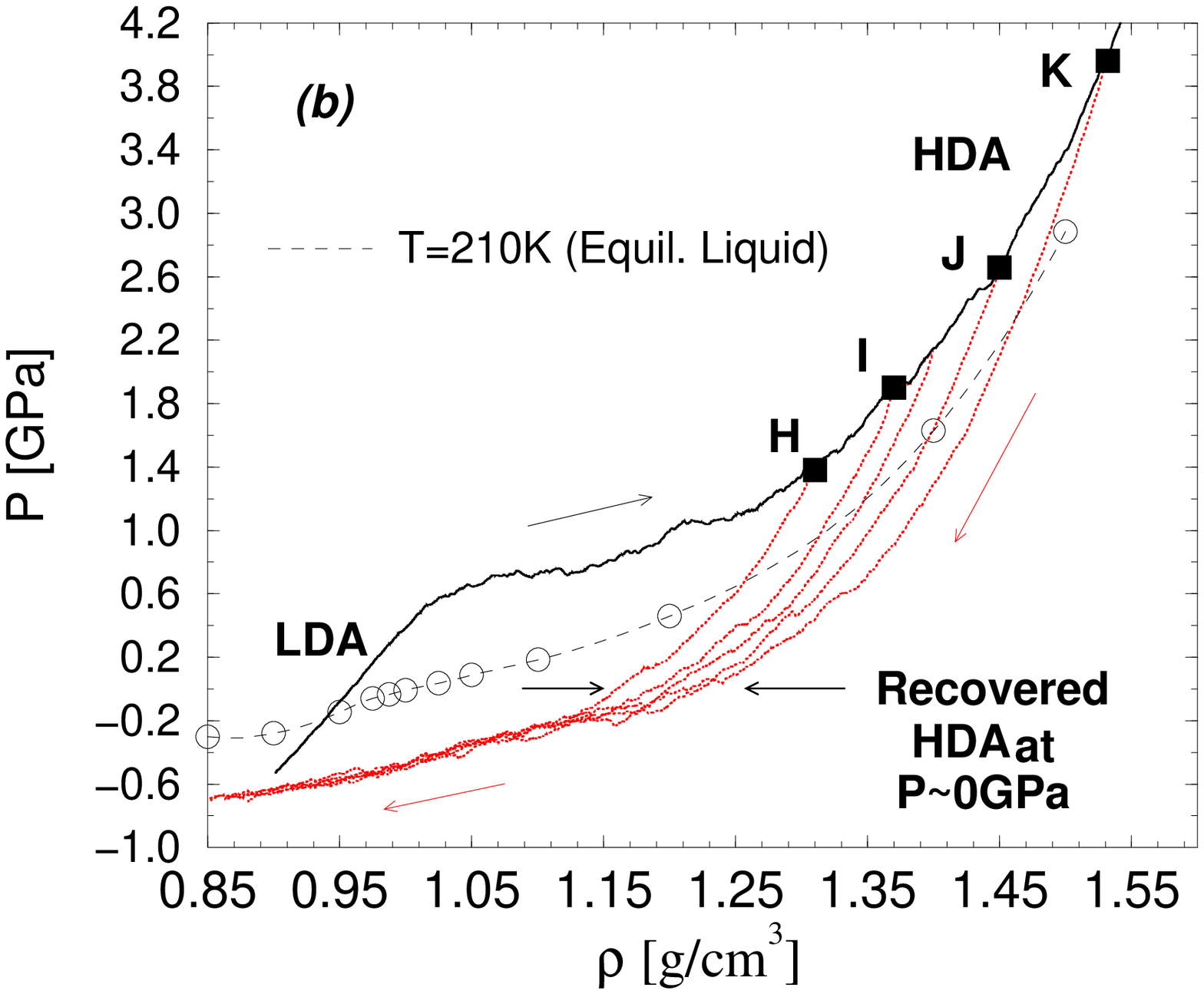}
  \hspace*{0.1cm}
  \epsfxsize=9cm
  \epsfbox{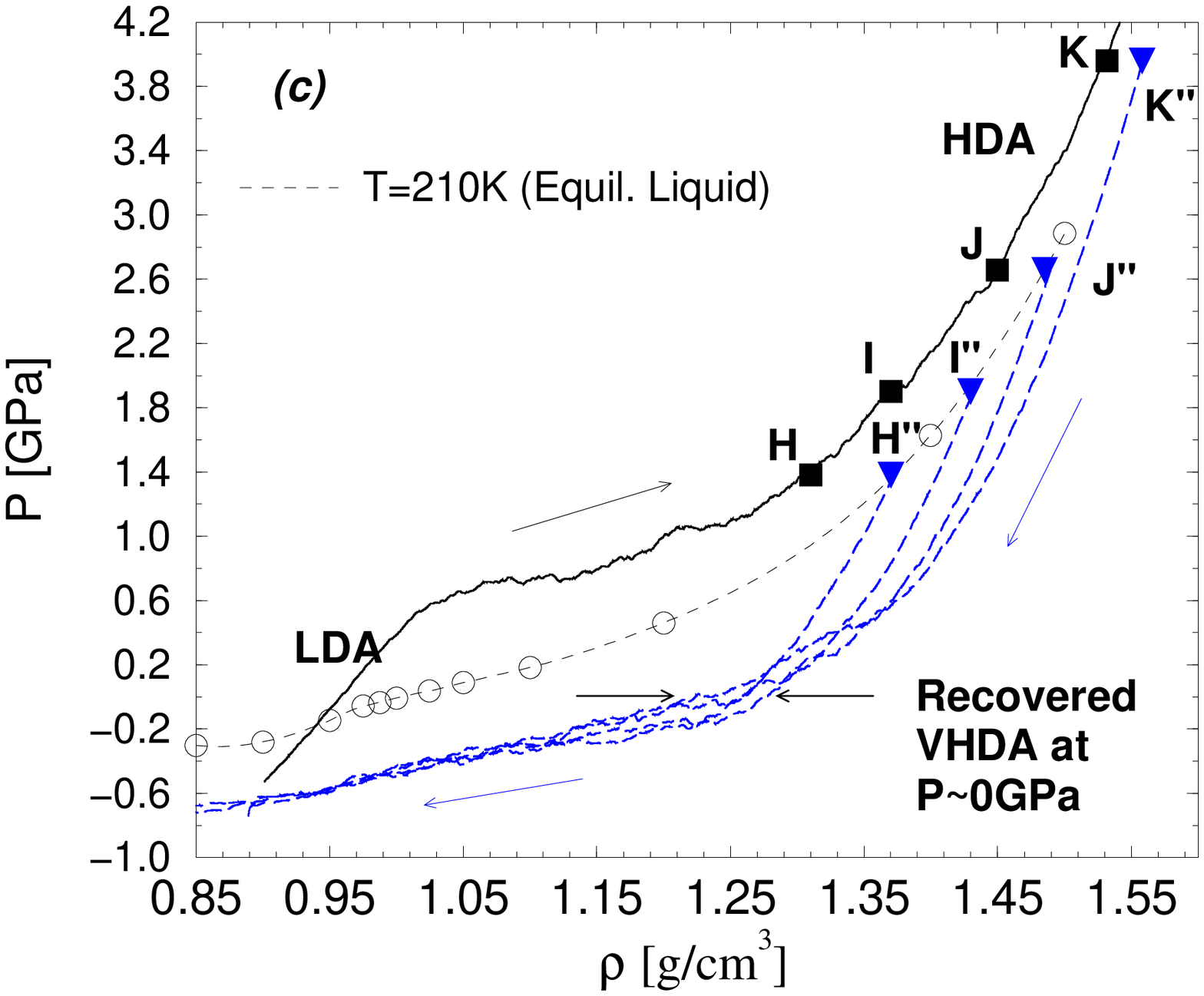}
}}

\vspace*{0.5cm}
\caption{(a) $P$ during decompression of the system starting at
  state points H, I, H'' and I''. Inset: State points H' and I' indicate
  VHDA configurations at $T\approx 165$~K obtained after heating the HDA configurations
  represented by state points H and I at $T=77$~K. State points H'' and I''
  correspond to the final VHDA states after cooling VHDA back to $77$~K.
  Decompressions at $T=77$~K of (b) the four HDA states indicated by state points (H, I,
  J, K) and (c) the four VHDA states indicated by (H'', I'', J'', K'').
  At $P \approx 0$~GPa, the recovered configurations of HDA and VHDA do
  not collapse into two single states but they cover two ranges of
  densities.  HDA densities fall in the interval $1.15-1.24$~g/cm$^3$
  while VHDA densities fall in the interval $1.22-1.28$~g/cm$^3$.  All
  recovered HDA and VHDA forms collapse to a single state at
  $\rho=1.05$~g/cm$^3$ and $P \approx -0.4$~GPa.}
\label{recoveredHDAVHDA}
\end{figure}

We first discuss state points H and I, the cases of HDA at  $P=1.38$~GPa and $P=1.9$~GPa [see inset of
Fig.~\ref{recoveredHDAVHDA}(a)]. State points H' and I' are the glasses
obtained after heating HDA, i.e., VHDA ($\approx 165$~K).  In agreement with
experiments, cooling VHDA back to $T=77$~K increases $\rho$.  The
final glasses after cooling, i.e., VHDA ($\approx 77$~K), are indicated
by state points H'' and I'' in the figure.  We note that VHDA($\approx 77$~K)
can be cycled back upon heating to VHDA ($\approx 165$~K).  This is also in agreement with the experimental results\cite{loerting}.
Figure~\ref{recoveredHDAVHDA}(a) shows the evolution of $P$ as a function of $\rho$ during decompression of the system starting at state points
H, I (for HDA), H'' and I'' (for VHDA). During decompression, VHDA is denser than the
corresponding HDA, and only at $P \approx -0.15$~GPa the densities  of the decompressed samples of VHDA are equal.

In experiments, HDA and VHDA recovered at  ambient $P$
and $T=77$~K
are characterized by only one state with $\rho=1.17$~ g/cm$^3$ and $\rho=1.25$~ g/cm$^3$, respectively. In Fig.~\ref{recoveredHDAVHDA}(b)
and~\ref{recoveredHDAVHDA}(c) we show the $P$ during decompression
of the four HDA and the four VHDA forms, respectively.  The recovered HDA
at $P=0$~GPa and $T=77$~K have different densities,  and do not collapse to a single state.  Also the recovered VHDA forms are characterized by different densities, notwithstanding the same $P$ and $T$. We observe that at $P \approx 0$~GPa, HDA densities fall in the interval $1.15-1.24$~g/cm$^3$ while VHDA densities fall in the interval $1.22-1.28$~g/cm$^3$. Furthermore, these two
$\rho$ intervals overlap, meaning that recovered HDA and VHDA are 
indistinguishable by their $\rho$ at zero $P$. 
  It is also clear from our results that there is a continuum of glasses
 at normal $P$. A continuum of glasses at normal $P$ has been already
 suggested by experiments in Refs.~\cite{science,urquidi2} and obtained in
 computer simulations using the TIP4P model \cite{parrinello}.  Our 
simulations also show that as $P$ decreases to negative values, all the
recovered HDA and VHDA forms collapse to a single state at 
$\rho=1.05$~g/cm$^3$ and $P \approx -0.4$~GPa.

\section{Radial Distribution Functions}
\label{rdfs}

\subsection{LDA and HDA at normal $P$}

\begin{figure}[p]
\narrowtext 

\centerline{
\hbox {
  \vspace*{0.5cm}  
  \epsfxsize=9cm
  \epsfbox{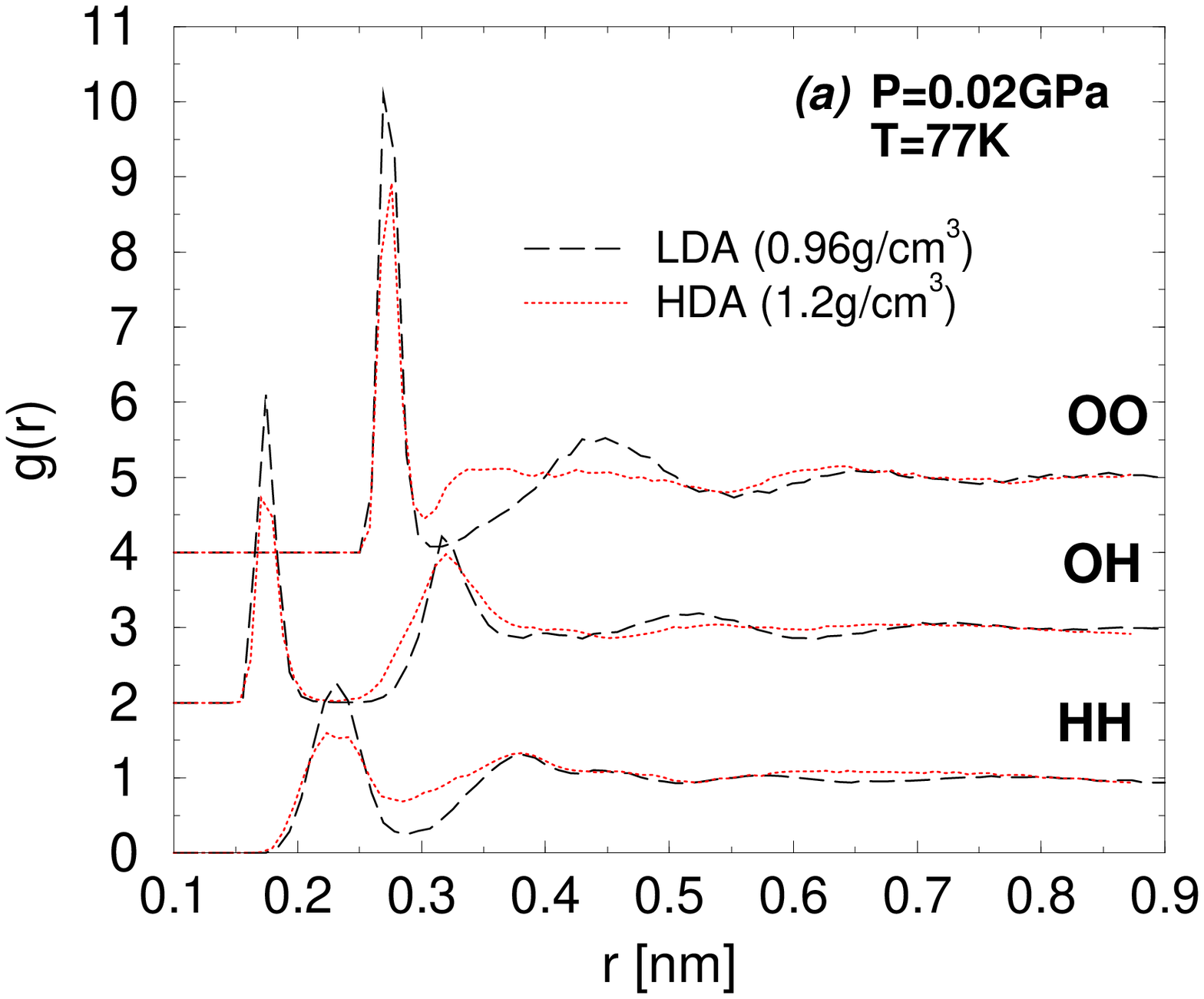}
}}
\centerline{
\hbox {
  \vspace*{0.5cm}  
  \epsfxsize=9cm
  \epsfbox{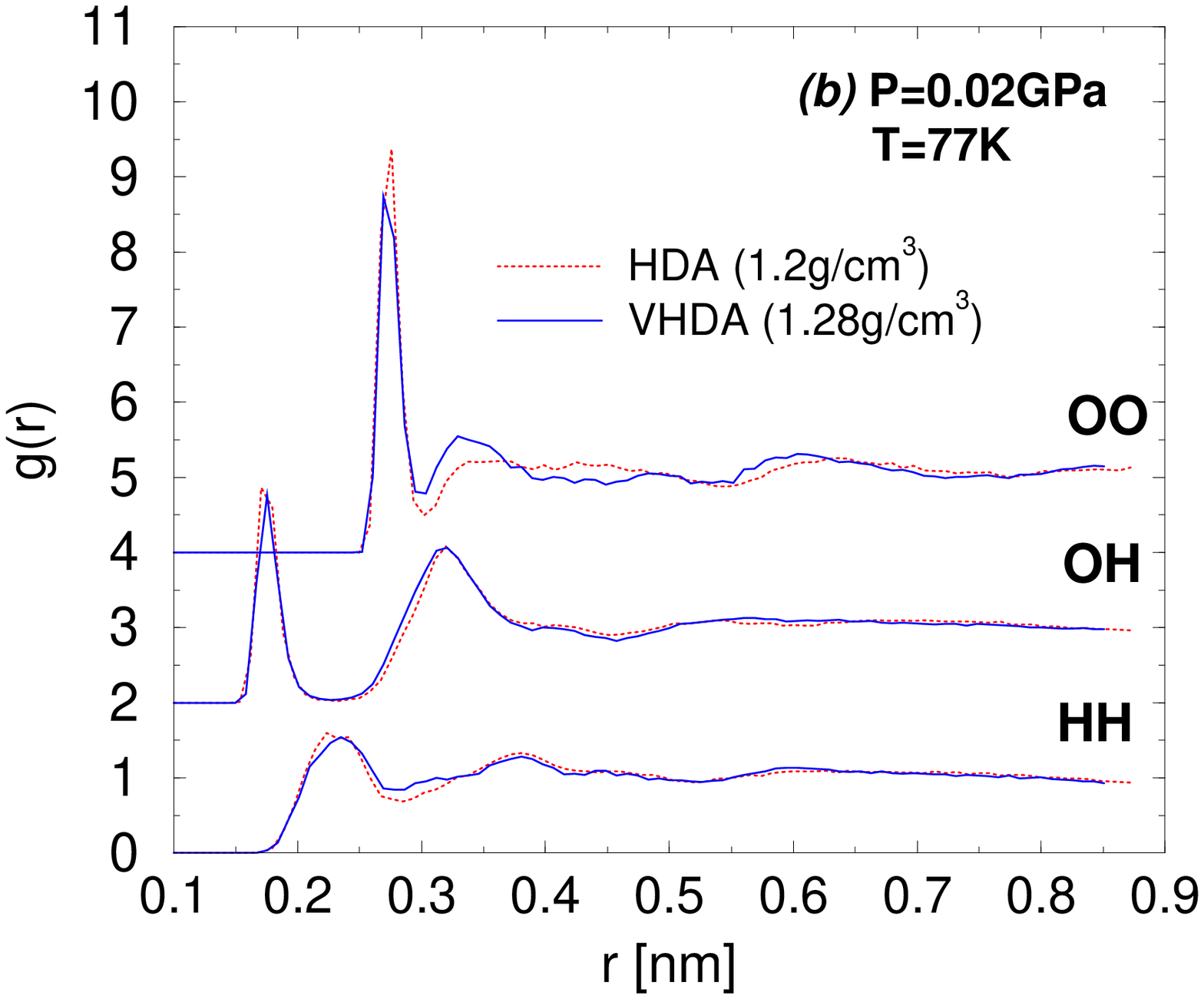}
}}
\vspace*{0.5cm}
\caption{Oxygen-oxygen ($OO$), oxygen-hydrogen ($OH$), and hydrogen-hydrogen
  ($HH$)  radial distribution functions at $=77$~K and $P=0.02$~GPa
 for (a) LDA and HDA, and (b) HDA and VHDA. The area under the RDFs indicate
 that LDA, HDA and VHDA are characterized by a tetrahedral hydrogen-bonded
 network  and that, in comparison with LDA, HDA has an extra interstitial
 neighbor between the first and second shell while VHDA has two extra
 molecules. 
}
\label{rdfLDAHDAVHDA-1atm}
\end{figure}

In this section, we study in detail the radial distribution function (RDF) of amorphous ice. We compare the results from 
simulations and experiments on LDA, HDA and VHDA. We also investigate how the structure of HDA
changes as HDA is transformed to VHDA at high $P$. The RDFs for
glassy water are commonly measured at $T \approx 80$~K and ambient
$P$.  In Fig.~\ref{rdfLDAHDAVHDA-1atm}(a) we show 
 the RDF obtained from simulations
 for LDA ($\rho=0.96$~g/cm$^3$) and recovered HDA 
($\rho = 1.2$~g/cm$^3$) both at $P=0.02\pm 0.02$~GPa and $T=77$~K.
  Recovered HDA is obtained by isothermal decompression at $T=77$~K of HDA
($\rho=1.37$~g/cm$^3$) [state point H in
Fig.~\ref{recoveredHDAVHDA}(a)], as discussed in the section above.
Figure~\ref{rdfLDAHDAVHDA-1atm}(a) can be compared with Fig.~1 in
Ref.~\cite{finneyLDAHDA}, which shows the experimental RDF of LDA and
recovered HDA at the state point ($P=1$~atm, $T=80$~K). Both figures are
very similar.  For example, the maxima of $g_{\mbox{\scriptsize
OO}}(r)$ for LDA in our simulations are located at $0.27$, $0.44$ and
$0.66$~nm, while experimentally \cite{finneyLDAHDA} they are located at
$0.27$, $0.45$ and $0.68$~nm. When going from LDA to HDA by compression,
the first peak of $g_{\mbox{\scriptsize OO}}(r)$ increases, the second
peak gets wider and smaller, and the third peak shifts to lower values
of $r$. However, for HDA we find some differences between experiments
and our simulations. The second peak of $g_{\mbox{\scriptsize OO}}(r)$
for our HDA simulations cannot be resolved into two peaks as clearly as
in experiments \cite{finneyLDAHDA}, and the first minimum in
$g_{\mbox{\scriptsize OO}}(r)$ is located at $r=0.30$~nm instead of at
$\approx 0.32$~nm. The same observations with respect to the second peak
in $g_{\mbox{\scriptsize OO}}(r)$ can be made from MD simulation
using the TIP4P model \cite{tse,parrinello}
 and from a reverse Monte Carlo simulation \cite{laszlo}. We
note that the second peak of $g_{\mbox{\scriptsize OO}}(r)$ is very
sensitive to $T$; measurements at $P=0$~GPa, $T=100$~K, $\rho
=1.16$~g/cm$^3$ show that the second peak of $g_{\mbox{\scriptsize
OO}}(r)$ cannot be resolved into two peaks \cite{soperHDA100}.

The plots for $g_{\mbox{\scriptsize OH}}(r)$ for LDA and HDA 
in Figure~\ref{rdfLDAHDAVHDA-1atm}(a) show the
same features as those found experimentally in Ref.\cite{finneyLDAHDA}.  When going from
LDA to HDA, the first peak of $g_{\mbox{\scriptsize OH}}(r)$ increases,
the second peak gets wider, and the
third peak practically disappears.  However, we note that for both LDA
and HDA the relative heights of the first and second peaks of
$g_{\mbox{\scriptsize OH}}(r)$ in experiments and simulations are
different.  Experiments \cite{finneyLDAHDA} show that both peaks have
almost the same amplitude for LDA, while for HDA the {\it first} peak is
smaller than the second one. Instead, SPC/E shows a larger first peak in
$g_{\mbox{\scriptsize OH}}(r)$ for both LDA and HDA suggesting that the structure is more tetrahedral than in real water. In accord with
experiment \cite{finneyLDAHDA}, for the case of $g_{\mbox{\scriptsize
HH}}(r)$ we find an increase of the first peak and decrease of the first minimum of the RDF.

\subsection{HDA and VHDA at ambient $P$}

Figure~\ref{rdfLDAHDAVHDA-1atm}(b) shows the RDF of recovered HDA
($\rho \approx 1.37$~g/cm$^3$) and recovered VHDA ($\rho=1.28$~g/cm$^3$) generated
as discussed in the previous sections ($P=0.02\pm 0.02$, $T=77$~K). VHDA was obtained
by isobaric heating of HDA ($\rho \approx 1.37$g/cm$^3$) up to $T\approx 165$~K.
Figure~\ref{rdfLDAHDAVHDA-1atm}(b) can be compared with Fig.~1 in
Ref.~\cite{finneyHDAVHDA}, which shows the experimental RDF for
recovered HDA and VHDA at ($P=1$~atm, $T=80$~K).  In accord with
experiments, when going from HDA to VHDA, $g_{\mbox{\scriptsize OO}}(r)$
shows a weak decrease in the first peak and a shift of the third peak to
lower values of $r$. Furthermore, the second wide peak of
$g_{\mbox{\scriptsize OO}}(r)$ for HDA practically disappears at $r
\approx 0.43$~nm. In the experimental $g_{\mbox{\scriptsize OO}}(r)$,
the second peak in HDA merges into the first peak, filling the gap at
$r=0.33$~nm, the first minimum for HDA. As a consequence, the first peak
develops a shoulder with a second maximum at $r=0.35$~nm.  In our
simulations, this does not occur. Instead, the second maximum is located
at $r \approx 0.33-0.34$~nm and the first minimum in HDA is not filled,
i.e., we still can see a clear minimum in $g_{\mbox{\scriptsize OO}}(r)$
for VHDA. The presence of such a minimum
 in the $g_{\mbox{\scriptsize OO}}(r)$ for VHDA has been observed also in MD simulations using  
the TIP4P model \cite{parrinello}.
 When moving from HDA to VHDA, we find that
$g_{\mbox{\scriptsize OH}}(r)$ barely changes.  There is a weak shifting
of the second peak toward lower values of $r$ as it  is found experimentally
\cite{finneyHDAVHDA}, and the observed decrease of the first peak is
barely seen in our simulations.  We find that $g_{\mbox{\scriptsize
HH}}(r)$ also shows weak changes. As in \cite{finneyHDAVHDA}, we find a
decrease in the first peak and increase of the first minimum upon heating.

We expect to find differences in the RDF between experiments and
simulations. Properties for the SPC/E potential at a given 
$T$ correspond to those in experiments found at approximately $T'
\approx T+40$~K \cite{poole}. As we already mentioned, the
$g_{\mbox{\scriptsize OO}}(r)$ we find for HDA at $T=77$~K is more similar to that found in experiments at $T=100$~K \cite{soperHDA100} than the corresponding one at $T=80$~K~\cite{finneyLDAHDA}. Recovering VHDA at higher $T$ will allow neighbor oxygen atoms to come closer, thereby filling the gap at $r\approx0.3$~nm in $g_{\mbox{\scriptsize
OO}}(r)$. Therefore, differences between simulations and experiments in the RDF of VHDA are expected to be smaller at higher $T$.

\subsection{Isobaric Heating of HDA to VHDA at high-$P$}

\begin{figure}[p]
\narrowtext 

\centerline{
\hbox {
  \vspace*{0.5cm}  
  \epsfxsize=9cm
  \epsfbox{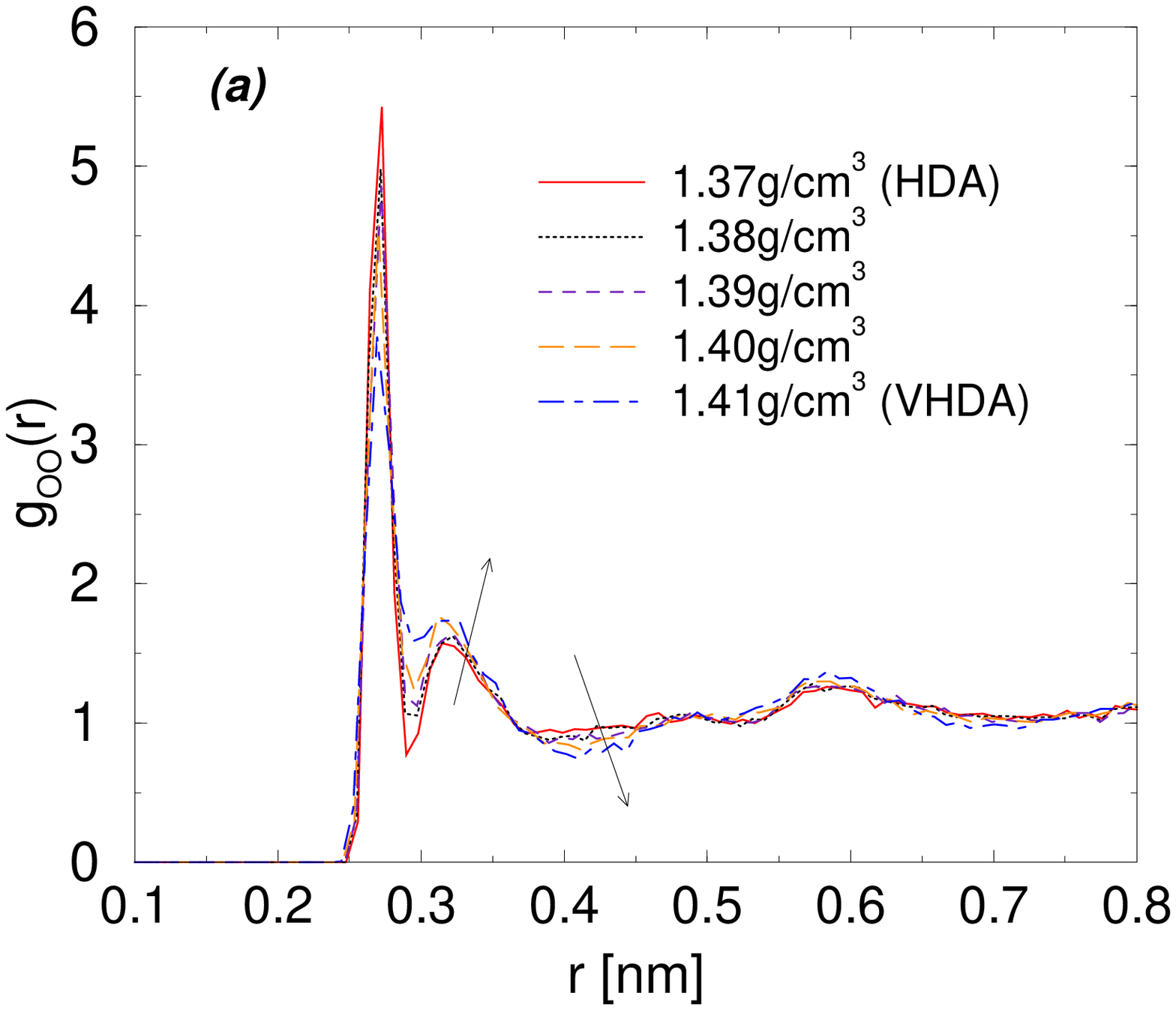}
  \hspace*{0.1cm}
  \epsfxsize=9cm
  \epsfbox{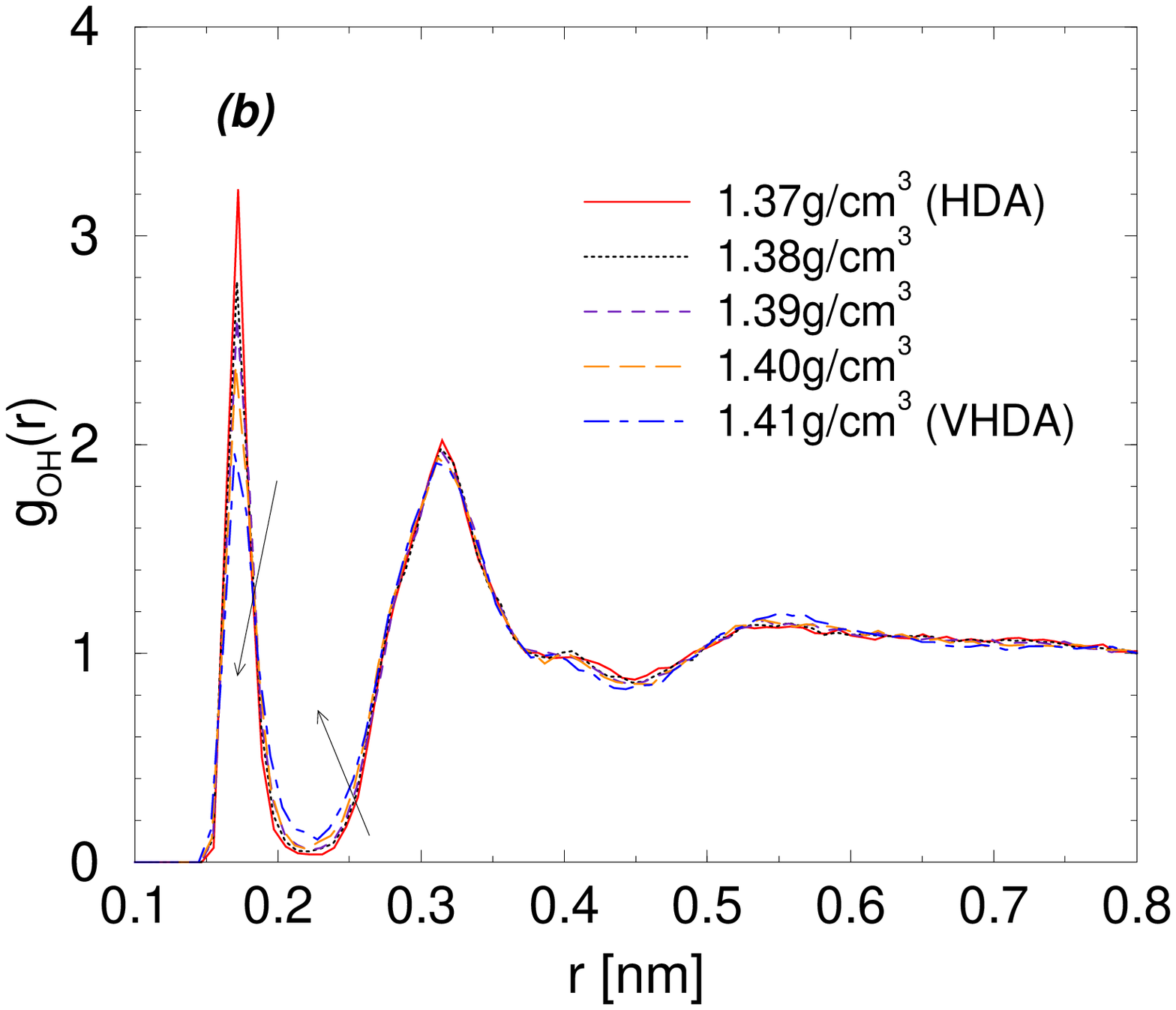}
}}
\centerline{
\hbox {
  \vspace*{0.5cm}  
  \epsfxsize=9cm
  \epsfbox{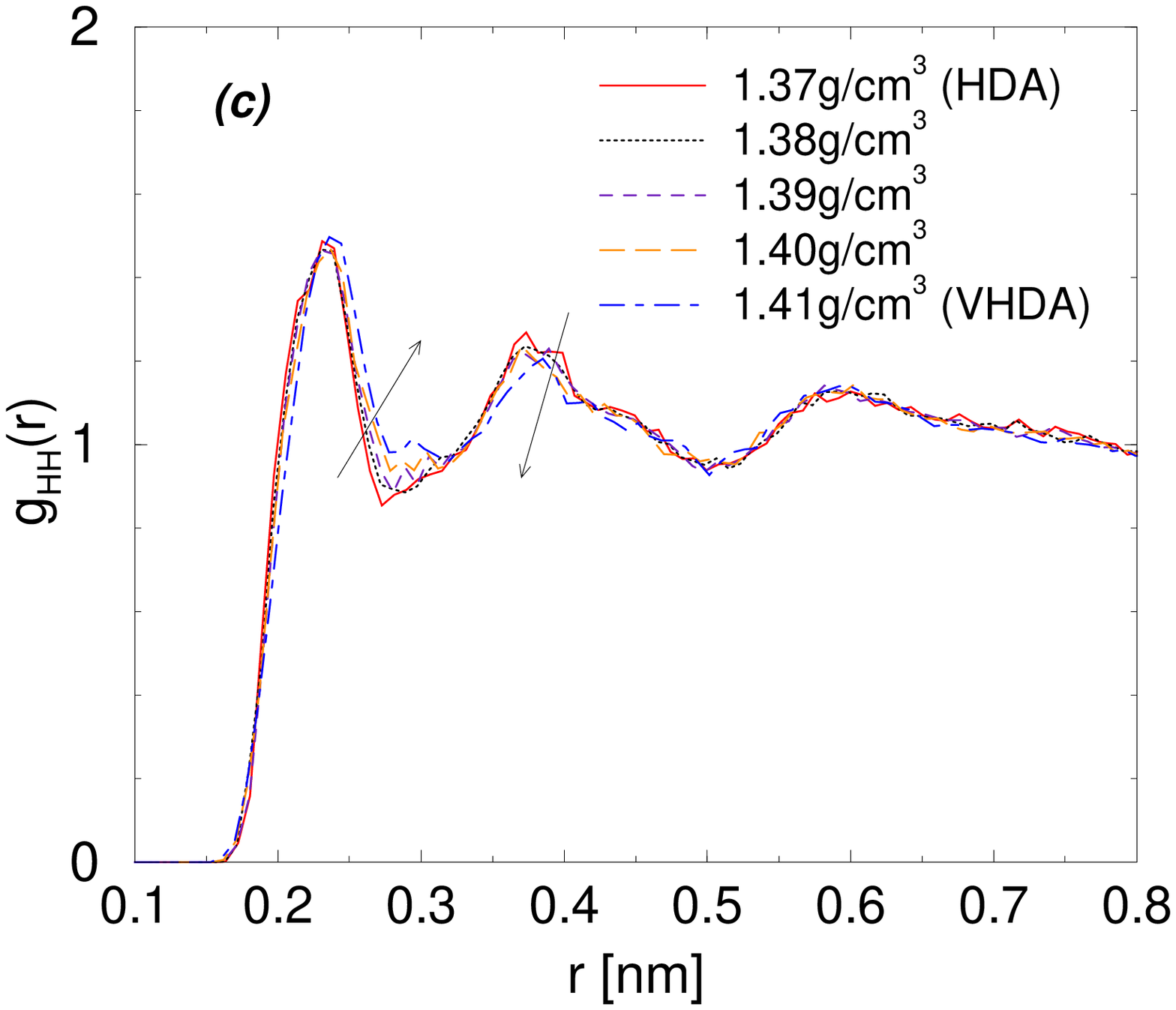}
}}
\vspace*{0.5cm}
\caption{Evolution of the radial distribution functions upon
 isobaric heating of HDA($1.37$~g/cm$^3$) at $P=1.9$~GPa from $T=77$~K
 up to $T \approx 175$~K.   }
\label{gr-HDAVHDA}
\end{figure}

In Fig.~\ref{gr-HDAVHDA} we show the evolution of the RDF on heating HDA ($\rho=1.37$~g/cm$^3$) at $P=1.9$~GPa from $T=77$~K up to $T \approx
175$~K.  Figure~\ref{gr-HDAVHDA}(a) shows $g_{\mbox{\scriptsize OO}}(r)$. Upon
heating HDA, the first peak in $g_{\mbox{\scriptsize OO}}(r)$ decreases and molecules move, filling the gap at $r=0.29$~nm. The second peak
shifts to lower values of $r$, merging with the first peak, and the
minimum at $r=0.41$~nm decreases. This RDF of VHDA is  remarkably similar to
the experimental distributions
 \cite{finneyHDAVHDA} for recovered VHDA at $P=0.02\pm 0.02$ and $T=77$~K,
suggesting that differences observed in our recovered configurations  set up during the decompression process. Figure~\ref{gr-HDAVHDA}(b) shows the evolution of
$g_{\mbox{\scriptsize OH}}(r)$ upon heating. The distribution shows a
decreasing first peak and increasing first minimum. These changes could not be observed in the recovered configurations of HDA and
VHDA at $P=0.02$~GPa. In Fig.~\ref{gr-HDAVHDA}(c) we note that
$g_{\mbox{\scriptsize HH}}(r)$ presents mainly a weak increase of the first minimum.

\subsection{Coordination Number of LDA, HDA and VHDA}

Integration of the experimental
$g_{\mbox{\scriptsize OH}}(r)$ at $T=80$~K and normal $P$ for  LDA, HDA, and
VHDA in the range $0.14$~nm~$<r<2.5$~nm
indicates that each oxygen ($O$) atom is surrounded by $2\pm0.1$ hydrogen
atoms \cite{finneyHDAVHDA,finneyLDAHDA}. Furthermore, integration of
$g_{\mbox{\scriptsize OO}}(r)$ for $r<3.1$~nm (the location of the first minimum in
the experimental VHDA $g_{\mbox{\scriptsize OO}}(r)$) indicates that each $O$ 
has $\approx 4$ nearest-neighbor $O$ atoms.  Therefore, LDA, HDA and VHDA are characterized by a tetrahedrally-coordinated fully hydrogen
bonded network.  

Our simulations agree with these experimental findings: we find from Fig.~\ref{rdfLDAHDAVHDA-1atm} that at $P=0.02$~GPa and $T=77$~K
 in LDA, HDA and VHDA each oxygen is surrounded by $1.95\pm0.5$ hydrogen atoms (for $r<2.5$~nm) and by $4-4.1$ oxygen atoms for $r<0.303$~nm (the first minimum of the simulated 
$g_{\mbox{\scriptsize OO}}(r)$ for VHDA).  Experiments  \cite{finneyHDAVHDA,finneyLDAHDA} also indicate that  differences in structure of HDA and VHDA as compared  to LDA basically arise from the presence of extra interstitial
neighbors. Integration of the experimental $g_{\mbox{\scriptsize OO}}(r)$ from
$r=0.31$~nm (the first minimum of $g_{\mbox{\scriptsize OO}}(r)$ for
VHDA) to $r=0.33$~nm (the first minimum of $g_{\mbox{\scriptsize
OO}}(r)$ for HDA), indicates that, in this range of distances,  an $O$ atom has $0.9$ ($\approx 1$) 
oxygen neighbors in HDA and $1.7$ ($\approx 2$) in VHDA. In  simulations, integrating from $r>0.303$~nm (first minimum of $g_{\mbox{\scriptsize OO}}(r)$ for VHDA) up to
$r=0.33$~nm, we find that each $O$ atom has $1.1$ ($\approx 1$)
neighbor $O$ in HDA and $1.8$ ($\approx 2$) in VHDA  (in LDA this value is only 
 $0.1$).  Therefore, simulations are consistent with  experiments showing 
that, in comparison to LDA, there is one extra interstitial molecule in
the structure of HDA and approximatively two in the structure of VHDA.

\section{The HDA-LDA, VHDA-LDA and VHDA-HDA transformations}
\label{x-x}

\subsection{HDA-LDA transformation}

It was first observed by Mishima et {\it al.} \cite{mishima84} that recovered HDA ($\rho \approx 1.17 \pm 0.02$~g/cm$^3$) at $T=77$~K and ambient
$P$ transforms to LDA ($\rho \approx 0.94 \pm 0.02$g/cm$^3$)
 upon isobaric heating to $T \sim 117$~K.
 In Ref.~\cite{tse} it is found that simulations at ambient
$P$ also show this transition between HDA and LDA. Here
we show that our simulations at ambient $P$ show remarkably similar results to those found in \cite{tse}. However, we point out that
a direct HDA$\rightarrow$LDA transition with the 
SPC/E model at the present heating rate $q_h=3 \times 10^{10}$~K/s
takes place only at negative $P$.

\begin{figure}[p]
\narrowtext 

\centerline{
\hbox {
  \vspace*{0.5cm}  
  \epsfxsize=9cm
  \epsfbox{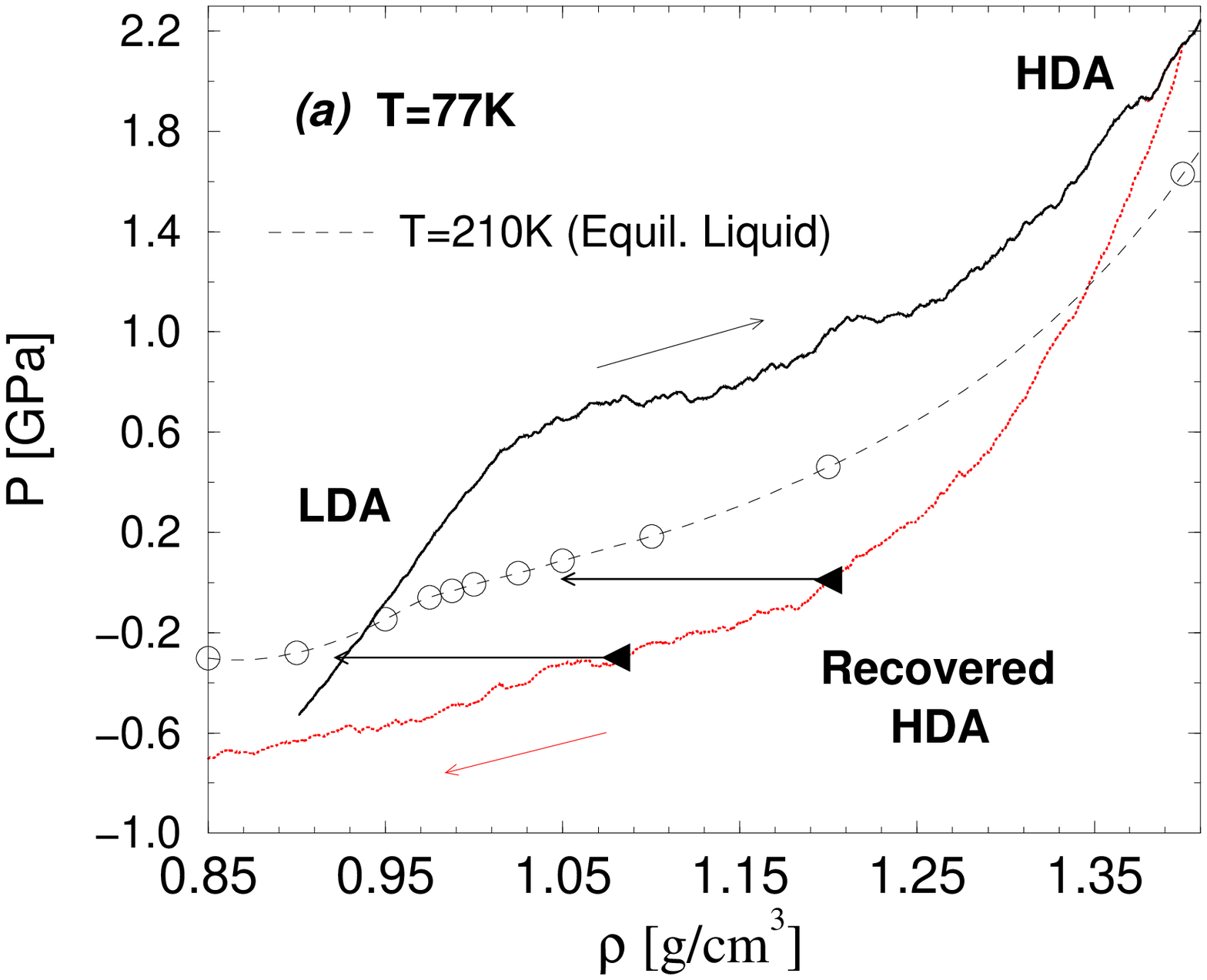}
  \hspace*{0.1cm}
  \epsfxsize=9cm
  \epsfbox{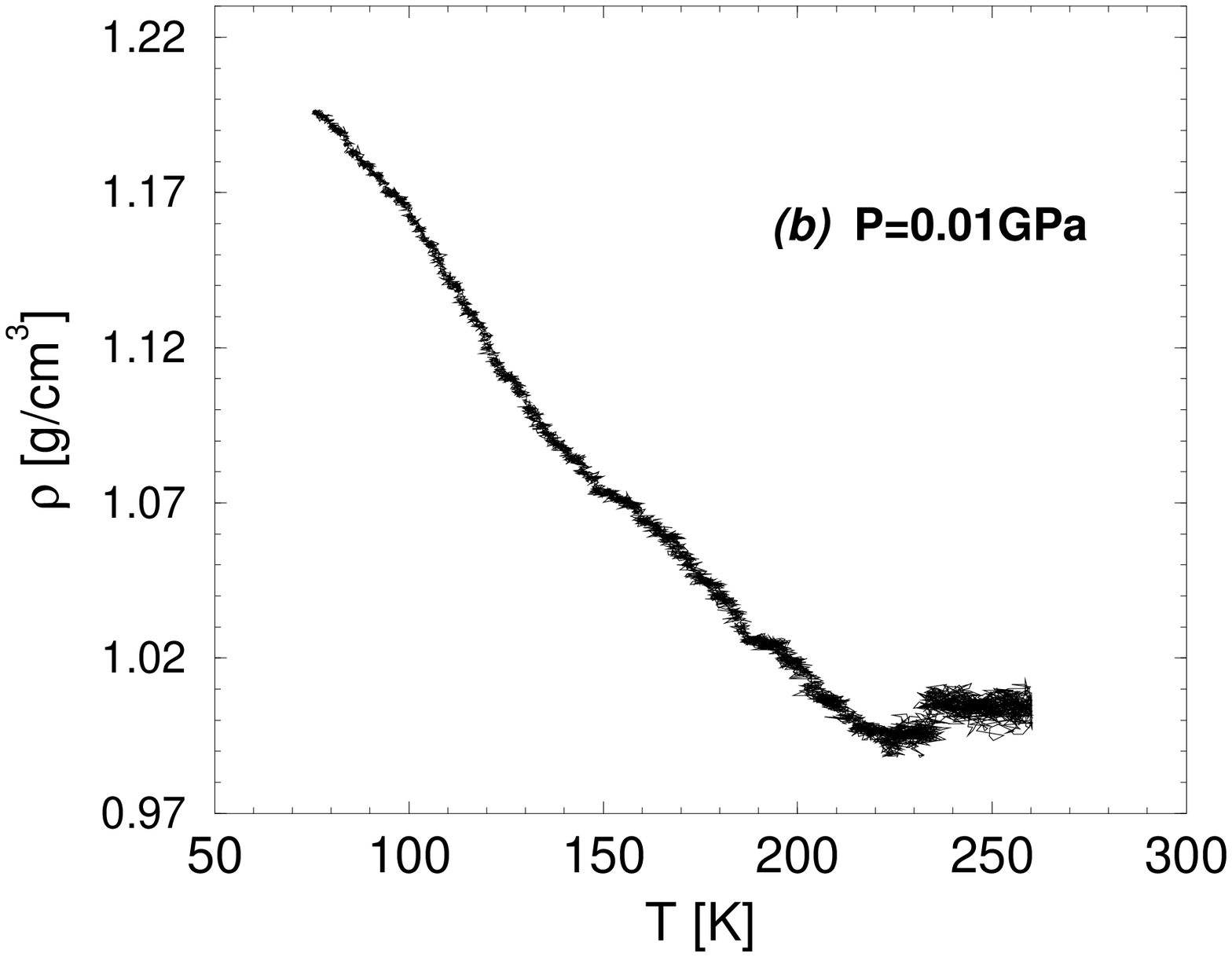}
}}
\centerline{
\hbox {
  \epsfxsize=9cm
  \epsfbox{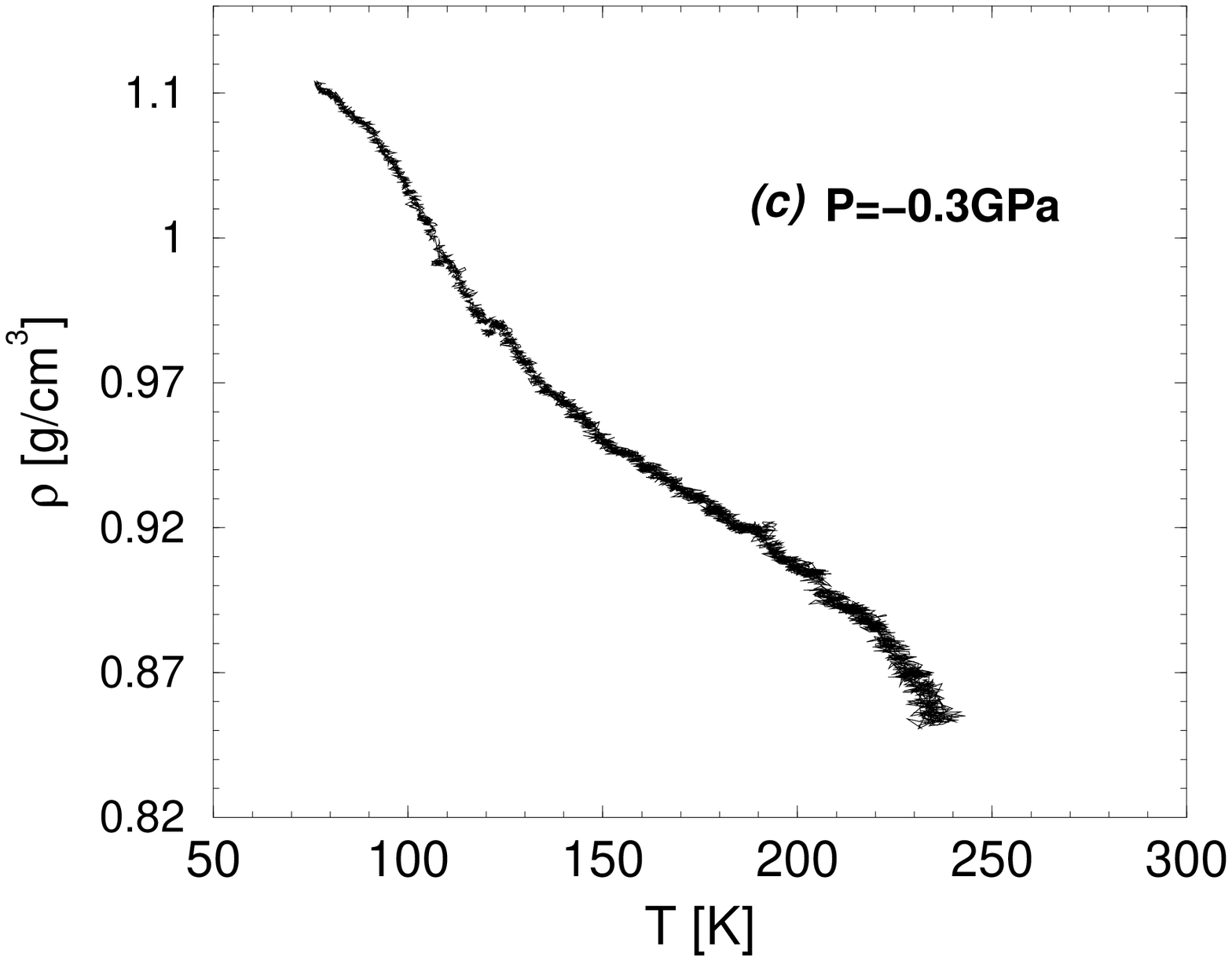}
}}

\vspace*{0.5cm}

\caption{(a) Compression of LDA to obtain HDA, and subsequent
  decompression of HDA from $1.4$~g/cm$^3$ down to $0.85$~g/cm$^3$.
  Triangles indicated the starting configurations of recovered HDA at
  ($1.2$~g/cm$^3$, $0.01$~GPa) and ($1.08$~g/cm$^3$, $0.01$~GPa) that
  are heated at constant $P$. The $\rho$ of the system upon heating
  is shown for (b) $P=0.01$~GPa and (c) $P=-0.3$~GPa.  In both cases,
  the system evolves toward the $T=210$~K liquid isotherm. Arrows in (a)
  indicate the $\rho$ change upon heating 
  up to $T=170$~K, before the liquid phase is reached.
 Only the isobaric heating at $P=-0.3$~GPa allows the system to reach the
  $\rho$ corresponding to LDA ($\approx 0.93$~g/cm$^3$ for this
  $P$) supporting the possibility of a HDA$\rightarrow$LDA
  transformation.    }
\label{hda_recoveredlda}
\end{figure}

Configurations of HDA ($\rho =1.4$~g/cm$^3$) are obtained by isothermal
compression of LDA at $77$~K up to $P=2.14$~GPa with a rate of 
$\partial \rho / \partial t = 5 \times 10^{-5}$~g/cm$^3$/ps, as explained in
 Sec.~\ref{simulations}. 
HDA is recovered by isothermal decompression at $T=77$~K  down to $P\approx
-0.7$~GPa with a rate of $\partial \rho
/ \partial t = -5 \times 10^{-5}$~g/cm$^3$/ps [see Fig.~\ref{hda_recoveredlda}(a)].  We then heat
isobarically recovered HDA ($\rho =1.2$~g/cm$^3$) at $P=0.01$~GPa.  The
evolution of the system in the $P$-$\rho$ plane is indicated by the upper
horizontal  arrow in Fig.~\ref{hda_recoveredlda}(a). $\rho(T)$  upon heating
at $P=0.01$~GPa is shown in Fig.~\ref{hda_recoveredlda}(b). 
We observe in Fig.~\ref{hda_recoveredlda}(b) that $\rho \approx 1.00$~g/cm$^3$ at
$T=220$~K, while the $\rho$ of LDA ($T=77$~K, $P=0.01$~GPa) is $\approx 0.96$~g/cm$^3$. 
Therefore, with the present rate $q_h=3 \times 10^{10}$~K/s,
 even when the system becomes less dense upon heating, it does not reach the
 $\rho$ corresponding to LDA at $77$~K and $0.01$~GPa.  Simulations using the TIP4P model are similar (Fig.~1 in Ref.~\cite{tse}). In Ref.~\cite{tse}, the $\rho$ of the system at $T
\approx 220$~K (during the heating process) is even larger than
$1$~g/cm$^3$. The $\rho$ corresponding to LDA in Ref.~\cite{tse} occurs when the $T$ is $T \gtrsim 260$~K and the system is in the liquid phase.

To find the HDA-LDA transition at $P=0.01$~GPa, the system has to ``move'' in Fig.~\ref{hda_recoveredlda}(a) (upper horizontal arrow)  from $\rho =1.2$~g/cm$^3$ to $\rho =0.96$~g/cm$^3$ crossing liquid isotherms,
like the  $210$~K-liquid isotherm shown in the figure.  
In the previous paragraph, we show that at the present $q_h$ the system gets
trapped at a state with $\rho$ corresponding to the liquid phase at
$T=210$~K. It is not clear whether a much slower $q_h$ could allow the system
to reach the $\rho$ of LDA. To find a direct HDA$\rightarrow$LDA
transformation with $q_h=3 \times 10^{10}$~K/s we heat recovered HDA at
negative $P$s. This is motivated because at $P<0$ the system should cross {\it
first\/} the glass isotherms {\it and then\/} the liquid isotherms to reach
the liquid phase. We confirm this view by heating isobarically recovered HDA at $P=-0.3$~GPa
[see lower horizontal arrow in Fig.~\ref{hda_recoveredlda}(a)]. $\rho(T)$
upon heating is shown in Fig.~\ref{hda_recoveredlda}(c) and is consistent
with the possibility of a HDA$\rightarrow$LDA transformation at $P=-0.3$~GPa, without any liquid intermediate state.

We note that we also repeated the isobaric heating procedure at $P=-0.71$~GPa, where ``recovered HDA'' (which is already LDA) has a $\rho$ of
$0.82$~g/cm$^3$. On isobaric heating $\rho$
decreases abruptly and the glass transforms to gas.  In this case, the system on the P-$\rho$ plane moves away from the LDA isotherms and the
glass eventually transforms into gas because there are no liquid isotherms at this $P$.

\subsection{VHDA-LDA transformation}

Experiments show that isobaric heating of ``recovered VHDA'' at $P= 0.11$~GPa from $T=77$~K to $T \approx 127$~K produces LDA \cite{loerting}.  We test this
experimental finding with our simulations by heating configurations of
VHDA at  $P= 0.11$~GPa. We generate VHDA configurations by
heating HDA ($\rho = 1.31$~g/cm$^3$, $P=1.38$~GPa) from $77$~K to
 $T \approx 160$~K.  We then recover VHDA at $T=77$~K, $P= 0.11$~GPa and
 $\rho = 1.27$~g/cm$^3$.
 
\begin{figure}[p]
\narrowtext 

\centerline{
\hbox {
  \vspace*{0.5cm}  
  \epsfxsize=8.5cm
  \epsfbox{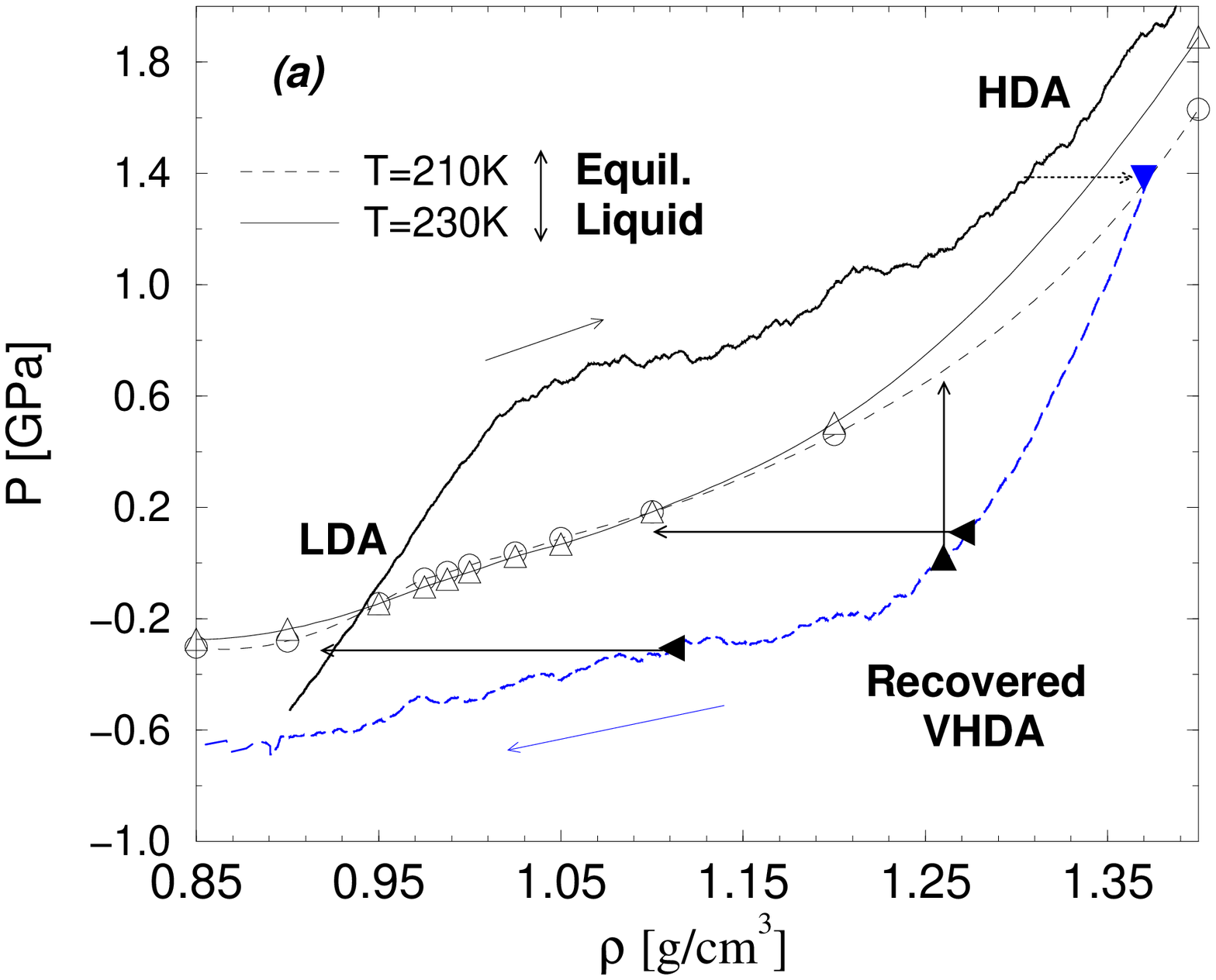}
  \hspace*{0.1cm}
  \epsfxsize=8.5cm
  \epsfbox{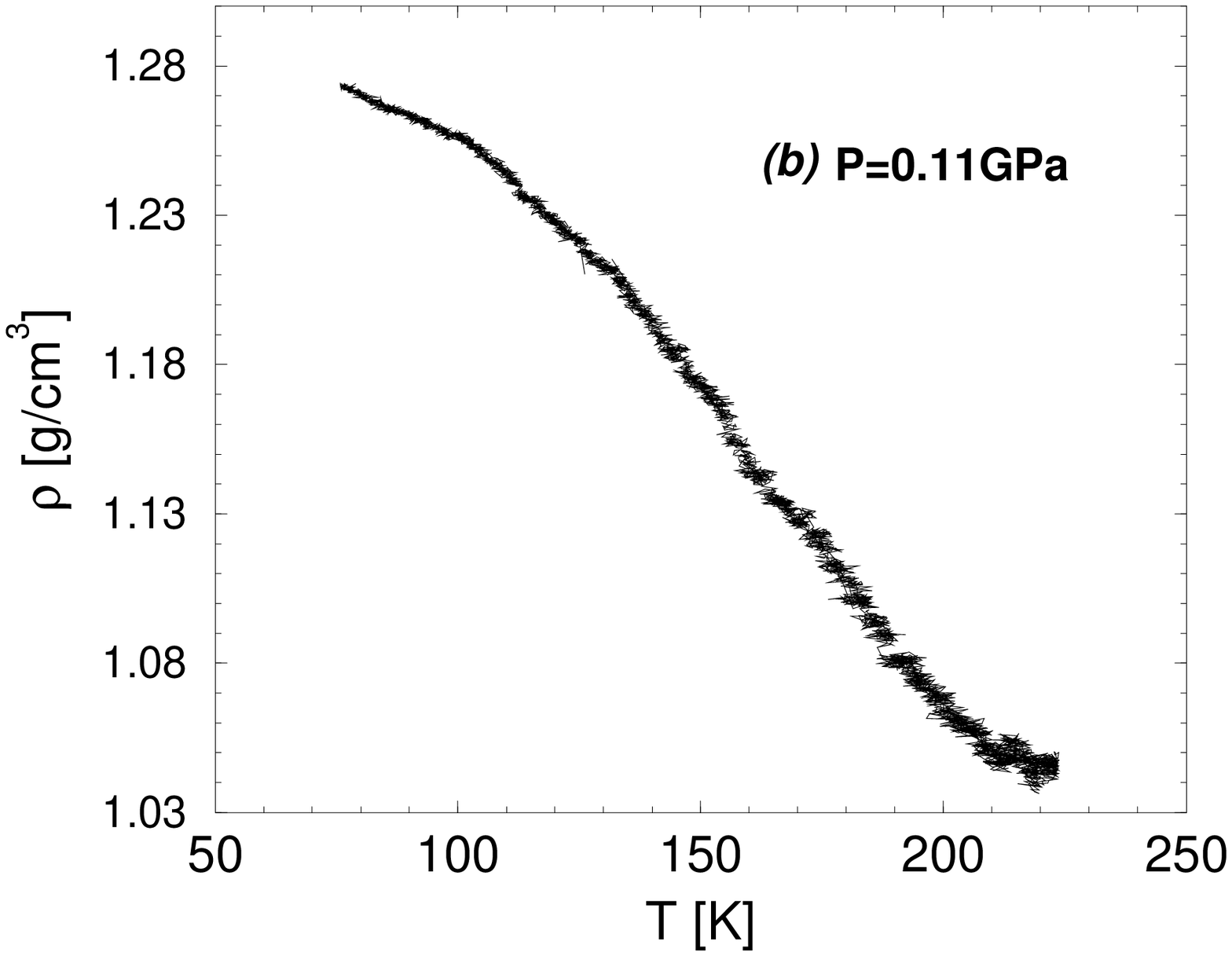}
}}
\centerline{
\hbox {
  \epsfxsize=8.5cm
  \epsfbox{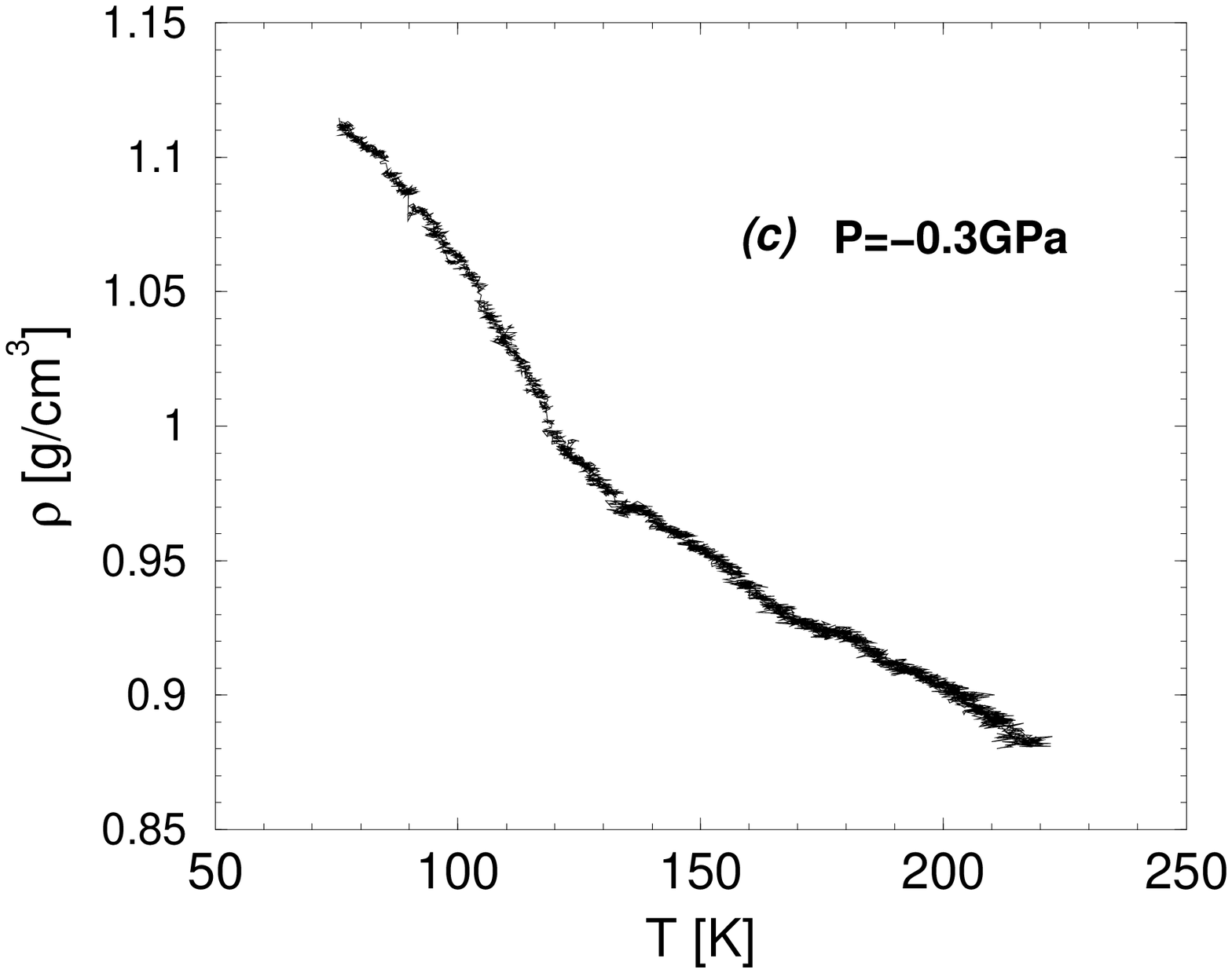}
  \hspace*{0.1cm}
  \epsfxsize=8.5cm
  \epsfbox{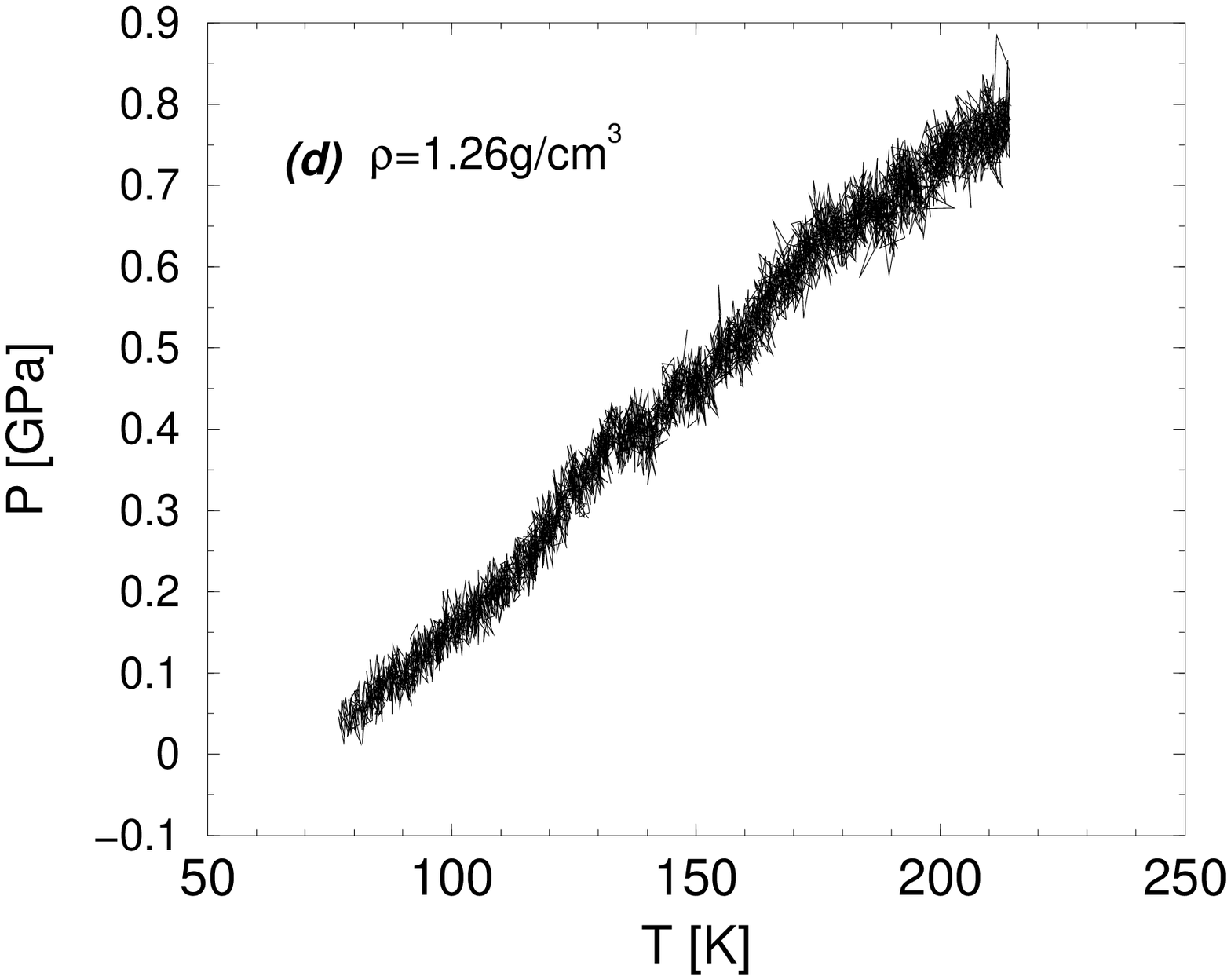}
}}

\vspace*{0.5cm}

\caption{(a) LDA ($\rho = 0.9$~g/cm$^3$) is compressed at $T=77$~K to obtain
  HDA  ($\rho \approx 1.31$~g/cm$^3$). Then, HDA is heated isobarically
  (horizontal 
  dotted arrow) to obtain VHDA (filled triangle down) and expanded at
  $T=77$~K (dashed line).  VHDA recovered at $\rho= 1.27$~g/cm$^3$
  (upper filled triangle left) is heated at constant $P=0.11$~GPa to
  test whether it produces LDA (see panel (b)). At this $P$, the
  system gets trapped in the liquid phase and no transformation to LDA
 is observed.  VHDA recovered at $\rho= 1.11$~g/cm$^3$ (lower
  filled triangle left) is heated at constant $P=-0.3$~GPa. 
The evolution of $\rho$ upon heating (see panel (c)) is consistent with 
the possibility of a VHDA$\rightarrow$LDA transformation.  VHDA
  recovered at $\rho= 1.26$~g/cm$^3$ (filled triangle up) is heated at
  constant $\rho=1.26$~GPa to test whether it produces HDA (see panel
  (d)).  The location of the $T=210$~K liquid isotherm avoids a
  VHDA$\rightarrow$HDA transformation (at the present heating rate $q_h=3
  \times 10^{10}$~K/s).}
\label{vhda_recovered-ldahda}
\end{figure}

Upon isobaric heating at $P=0.11$~GPa (see upper horizontal solid arrow
in Fig.~\ref{vhda_recovered-ldahda}(a)), the system expands as shown in
Fig.~\ref{vhda_recovered-ldahda}(b). However, the system does not reach
the LDA state. For this, the final $\rho$ before the glass transition $T$ should be $0.97$~g/cm$^3$, the $\rho$ of LDA [see
Fig.~\ref{vhda_recovered-ldahda}(a)].  Instead,
Fig.~\ref{vhda_recovered-ldahda}(b) shows that the $\rho$ at $T \approx
170$~K is $\approx 1.12$~g/cm$^3$, close to the $\rho$ of equilibrium
liquid at $T= 210$~K and $P= 0.11$~GPa---suggesting that the isobaric
heating of the recovered VHDA in our simulations brings the system to
the liquid state without an intermediate LDA. The situation is similar to the above discussion of the HDA$\rightarrow$LDA transformation at $P=
0.11$~GPa.  The glass upon heating evolves to the liquid phase with no intermediate transformation to LDA. In analogy with the situation of the
HDA$\rightarrow$LDA transition, a VHDA$\rightarrow$LDA transition could
 be obtained, with the present heating rates, at negative $P$s, i.e., at $P$ such that the system crosses {\it first\/} the glass isotherms {\it and then\/} the liquid isotherms.
The $\rho$ upon isobaric heating at $P=-0.3$~GPa of recovered VHDA is shown in
 Fig.~\ref{vhda_recovered-ldahda}(c) [see also the lower horizontal arrow in Fig.~\ref{vhda_recovered-ldahda}(a)] and it is consistent with the possibility of a VHDA$\rightarrow$LDA transformation.

\subsection{VHDA-HDA transformation}

Experiments show that isochoric heating of recovered VHDA at $P=0.02$~GPa and $T=77$~K up to $T=140$~K appears to produce HDA \cite{loerting}.  In fact, the x-ray
pattern of the final sample has been interpreted as a mixture of VHDA and HDA. We use our simulations to test for the presence of  VHDA$\rightarrow$HDA transformation by heating configurations of VHDA at
constant volume.  Configurations of VHDA are obtained by isobaric heating of HDA ($77$~K, $1.31$~g/cm$^3$) at $P=1.38$~GPa up to $T
\approx 160$~K. VHDA is then recovered at ($P=0.02$~GPa, $T=77$~K, $\rho=1.26$~g/cm$^3$).  The evolution of the system in the $P-\rho$
plane during isochoric heating at $\rho=1.26$~g/cm$^3$ from $T=77$~K up
to $T \sim 170$~K is indicated by the vertical arrow in Fig.~\ref{vhda_recovered-ldahda}(a).  $P(T)$ is shown in Fig.~\ref{vhda_recovered-ldahda}(d).  $P$
increases monotonically upon heating and approaches the value $0.72$~GPa
at $T \approx 200$~K.  At this $P$ and $T$,
the same thermodynamic properties as the
equilibrium
liquid at $T=210$~K are recovered, as indicated by the liquid isotherm in Fig.~\ref{vhda_recovered-ldahda}(a).  Therefore, our simulations do not
show a VHDA-HDA transformation because the system melts in the
liquid phase. The situation is similar to the
HDA$\rightarrow$LDA transformation at $P=0.01$~GPa discussed above.
The  $P$ increase during  isochoric heating  is a consequence of the VHDA$\rightarrow$liquid transition.

It is not clear whether a much slower heating rate (nowadays, not  accessible in  computer simulations\cite{french}) will show a VHDA$\rightarrow$HDA transformation or whether such a transformation does not exist. In the short term, experiments may help clarify this issue.

\section{Discussion}
\label{summary}

The present study was inspired by several motivations. Among them were (i) the
goal of testing how closely simulations based on the SPC/E potential reproduce
the experiments, and (ii) the goal of deriving a concise view of the phase
diagram of amorphous ice. 

We find that even using the simple SPC/E model for water,
simulations of isothermal compression of LDA produce another glass, HDA. 
By repeating the process at different compression $T$ we 
find that the glass isotherms cross one another in the $P$-$\rho$ plane.
This leads to the different behavior of glassy water at low and high $P$.

We investigate how under isobaric heating HDA converts into a denser glass,
 VHDA, while  under isobaric heating LDA evolves toward a less dense state.
 The higher is the compression $T$ of LDA, the higher is the $\rho$  of the final
 HDA form at a given $P$, and the closer it is to the corresponding VHDA. This
 finding is consistent with  the view of VHDA as the result of relaxation of
 HDA upon isobaric heating---i.e., the higher the compression $T$, the more the glass is
 able to relax during compression, approaching VHDA. In fact, by aging HDA
at high $P$ before heating to obtain VHDA we observe an increase  of $\rho$
 with time suggesting that VHDA is the result of relaxation of HDA. A similar
 interpretation of HDA as an unrelaxed glass was proposed by Mishima 
when forming HDA by compression of ice $I_h$ at different $T$
 \cite{mishima96}.

We also investigate the thermal stability of the glasses formed during the
compression process of LDA. To do this, we repeat the protocol to transform
HDA into VHDA at high-$P$. 
In other words, we perform isobaric heatings of samples of LDA and samples
obtained during the transformation of LDA to HDA.  As $T$ increases, 
irreversible relaxation processes take place generating annealed
versions of the starting samples. The resulting annealed glasses (obtained at
low and intermediate $P$) are analogous to the VHDA (obtained at high $P$).

Densities at different $P$s and $T$s for LDA, HDA and VHDA 
 agree well with experiments. Furthermore, comparison of the structure of
 these glasses at the level of the radial distribution functions
 shows differences between the three forms---LDA, HDA, and VHDA---similar to
 those observed in experiments. In particular, we find that
LDA, HDA and VHDA are characterized
 by a tetrahedral hydrogen bonded network. As compared to LDA, HDA has an
 extra interstitial neighbor between the first and second shell while VHDA
 has two extra molecules. 

Stimulated by experimental work,
  we also simulate the recovering process of different samples
 of HDA and VHDA at normal $P$. We find a continuum range of densities for
  each family of glasses. Furthermore, the ranges of densities of recovered
  HDA and recovered VHDA overlap at normal $P$. A continuum of glasses was
  proposed in Ref.~\cite{science} and  mentioned as a possibility in
  Ref.~\cite{loerting}. 

We discuss the LDA, HDA and VHDA 
results with respect to the equation of state of
liquid water. The comparison of equilibrium liquid
 isotherms with those for the glassy state in our simulations allow one to
predict the effect of cooling/heating and compression/decompression of
amorphous and/or liquid water.  For example, one can understand why it is
possible to find HDA$\rightarrow$LDA and VHDA$\rightarrow$LDA
transformations in SPC/E at $P=-0.3$~GPa and why it is not the case  at
$P=0.01$~GPa (at least with the present heating rate).
 We note that experimentally these two transformations
to LDA occur at $P\approx 0$~GPa. However, we expect
shifts in the values of $P$, $T$ and $\rho$ when comparing experiments and
simulations.  The possible VHDA$\rightarrow$HDA
transformation\cite{loerting} is not observable in simulations with the SPC/E
model with the present heating rate. 

Some experiments and computer simulations performed in the last few years  have challenged the phase diagram proposed in \cite{geneNature,poolepre,poole}.
Aging effects and the previous history of the glass (especially at $T
\lesssim T_g(P)$) make it difficult  to characterize the glass state in an unique way. 
In fact, a glass depends on how it is prepared: cooling/heating rates,
compression/decompression rates and aging times are variables that have to be
included when describing the glass state \cite{ourPRE-RC,johari04}.
Therefore, a phase diagram for a metastable system has to be viewed  with
caution. In the phase diagram proposed in \cite{geneNature,poolepre,poole},
 a first order transition line separates LDA and HDA at low-$T$,
and continues into the liquid phase at high-$T$ separating the two liquid
phases, LDL and HDL. Furthermore, it has been hypothesized that this first
order transition line ends in a second (metastable) critical point
\cite{poolepre,poole,PELWater,masako,scalaStarr}. The key points in the 
 phase diagram proposed in \cite{geneNature,poolepre,poole} are
 that LDA is the glass obtained upon cooling of LDL while HDA is the glass  obtained upon cooling of HDL. Recent results and simulations call for a
 refinement of this picture.  In \cite{ourVHDA}, we show that VHDA,
 and not HDA, is the glass obtained upon cooling of HDL.   
  In this modified phase diagram, the first order transition line between LDL and HDL is
 continued in the glassy phase separating LDA and VHDA.  We note that the
 properties of the glass obtained upon cooling a liquid  can change on the
 cooling rate \cite{ourPRE-RC}. Therefore, VHDA and LDA  represent each a
 family of glasses, all of them obtained upon cooling (with 
 different cooling rates) of HDL and LDL, respectively.  Other unrelaxed
 glasses, e.g. HDA (and probably many other glasses
 \cite{science} which might depend, for instance, on aging)
 cannot be shown in a single phase diagram.  The phase diagram in
 \cite{geneNature,poolepre,poole} with a 
{\it first order transition line} 
identifies two distinct  liquid phases and the two families of glasses
associated to them.

\subsubsection*{Acknowledgments}

We thank C.A. Angell, A. Geiger, I. Kohl, M.M. Koza, P.H. Poole, A.K. Soper,
 F.W. Starr, and J. Urquidi for discussions, NSF grant CHE~0096892,
 and FIRB for support, and the Boston University Computation
 Center for allocation of CPU time. We also gratefully acknowledge the
 support of NSF through Collaborative
Research in Chemistry Grant No. CHE~0404699.

\begin{table}
\begin{tabular}{|c|c|c|}
Label & $\rho$ [g/cm$^3$] & P [GPa]\\ \hline
A & $0.90$ & $-0.55$\\
B & $0.94$ & $-0.17$\\
C & $0.96$ & $0.01$\\
D & $1.005$ & $0.41$\\
E & $1.065$ & $0.68$\\
F & $1.17$ & $0.84$\\
G & $1.255$ & $1.10$\\
H & $1.311$ & $1.38$\\
I & $1.375$ & $1.90$\\
J & $1.452$ & $2.66$\\
K & $1.532$ & $3.96$\\
\end{tabular}
\caption{Location in the P-$\rho$ plane of the starting configurations 
used during the isobaric heatings. State points A-D correspond to LDA
configurations while state points H-K represent HDA configurations. 
State points E-G correspond to configurations in the LDA-HDA transformation. 
}
\label{points} 
\end{table}

\newpage

\newpage

\newpage

\newpage

\newpage

\newpage

\newpage

\newpage

\newpage

\newpage

\newpage

\newpage

\newpage

\begin{references}
\bibitem{footnoteSize}
In much of the previous work on models for water,
 216-molecule systems have been employed.
Tests with larger systems have confirmed that size effects are
 numerically irrelevant. In MD simulation studies of glass-glass
 transition in water, we recall that the LDA-HDA transition has been
 investigated with a 216-molecule system  
{\protect \cite{poole,poolepre,ourLDAHDA}} 
  and in a 128-molecule system {\protect \cite{tse}},
 both showing good agreement with experiments.  We note that in 
Ref.~{\protect \cite{poole}} it is stated 
 that constant pressure simulations performed with the ST2 potential
 for water show that the $\rho=0.8$~g/cm$^3$ isochore 
obtained for a 1728-molecule system  can be reproduced with a 216-molecule
 system.
\end{references}
\end{document}